\documentclass{LMCS}

\usepackage{amsmath,amssymb,amsbsy,latexsym,qsymbols,stmaryrd}
\usepackage{mathpartir}
\usepackage[linktocpage]{hyperref}

\thicklines

\newtheorem{definition}{Definition}[section]
\newtheorem{theo}[definition]{Theorem}

\newtheorem{lemma}[definition]{Lemma}
\newtheorem{coro}[definition]{Corollary}

\newtheorem{example}{Example}

\newcommand{\maths}[1]{\newline\mbox{}\hfill$#1$\hfill\mbox{}\\[.2em]}

\def\Vert{\mspace{1 mu}|\mspace{1 mu}}

\newif\iflong \longtrue   
\newif\ifpopl \poplfalse   
\newif\ifREADERS \READERSfalse   

\newif\ifCONF \CONFfalse 

\newcommand{\reptext}{\ensuremath{\mathtt{Rep}}}
\newcommand{\replone}[1]{\ensuremath{\reptext_{#1}}}
\newcommand{\repl}[2]{\replone{#1}(#2)}

\newcommand{\capa}{\ensuremath{\mathsf{cap}}}

\newcommand{\Rcapar}[1]{\Ar{\langle\sf{#1}\rangle}}
\newcommand{\Rcap}{\Rcapar{\capa}}

\newcommand{\ds}{\mathsf{sd}}
\newcommand{\dd}{\mathsf{dd}}

\newcommand{\deep}{\mathsf{depth}}

\newcommand{\Pb}{\MAIF}

\newcommand{\true}{\mbox{\textsf{true}}}

\newcommand{\nil}{\ensuremath{\mathbf{0}}}

\newcommand{\fn}[1]{\mbox{$\mathrm{fn}(#1)$}}

\def\eqdef{\stackrel{\mathrm{{{def}}}}{=}}
\newcommand{\Rar}{\mbox{$\Longrightarrow$}}
\newcommand{\rar}{\mbox{$\longrightarrow$}}

\newcommand{\inname}{\ensuremath{\mathsf{in}}}
\newcommand{\outname}{\ensuremath{\mathsf{out}}}
\newcommand{\openname}{\ensuremath{\mathsf{open}}}
\newcommand{\inamb}[1]{\ensuremath{\mathsf{in~}#1}}
\newcommand{\outamb}[1]{\ensuremath{\mathsf{out~}#1}}
\newcommand{\openamb}[1]{\ensuremath{\mathsf{open~}#1}}
\newcommand{\amb}[2]{\ensuremath{#1[#2]}}
\newcommand{\mess}[1]{\ensuremath{\langle#1\rangle}}

\newcommand{\sat}{\ensuremath{\models}}
\newcommand{\eqL}{\mbox{$=_L$}}

\newcommand{\AAA}{\mbox{$\mathcal{A}$}}
\newcommand{\BBB}{\mbox{$\mathcal{B}$}}

\newcommand{\FFF}{\mbox{$\mathcal{F}$}}

\newcommand{\ltrue}{\mbox{\textbf{$\top$}}}
\newcommand{\lfalse}{\mbox{\textbf{$\bot$}}}

\newcommand{\rtr}{\mbox{$\triangleright$}}
\newcommand{\brtr}{\mbox{$\blacktriangleright$}}
\newcommand{\sometime}{\mbox{$\Diamond$}}

\newcommand{\always}{\ensuremath{\Box}}

\newcommand{\thereis}[2]{\exists \; #1 \, . \, #2} 

\def\defiDS{\!\stackrel{\mathrm{{{def}}}}{=}\!}

\newcommand{\satDS}[2]{#1 \models #2 }

\newcommand{\orr}{ \vee } 
\newcommand{\andd}{ \wedge} 
\newcommand{\zero}{{\mathbf{0}}} 
\newcommand{\all}[2]{\forall \; #1 \, . \, #2} 

\newcommand{\at}[2]{ #1 @ #2} 

\newcommand{\limp}{\triangleright}  
\newcommand{\limpE}{\brtr}

\newcommand{\myneg}{\neg\,} 

\newcommand{\tkpPOPL}{3pt}  
\newcommand{\tkpP}{15pt}  

\def\trans#1{\mbox{{\sc{\tt #1}}}}   

\newcommand{\Mybar}
{
{
\vskip .3cm 
 \rule{\hsize}{0.3mm} \vskip .7cm 
}
}

\newcommand{\shortaxiomC}[2]{
 \displaystyle{ \over #1} \; \trans{#2} }

\newcommand{\shortinfruleN}[3]{  
\displaystyle{#1 \over #2} {\; \trans{#3}}}

\def\sub#1#2{\ensuremath{\{#1\!/\!#2\}}}

\newcommand{\andalso}{\quad\quad}  

\def\abs#1#2{(#1)\: #2}                     

\newcommand{\msg}[1]{\langle{#1}\rangle } 

\newcommand{\tkp}{2pt}  

\newcommand{\mboxB}[1]{\mbox{\it (#1)}}
\def\midd{\vert} 
\newcommand{\equivE}{\mathrel{\equiv_{\rm E}}}  

\newcommand{\MA}{\mbox{MA}}
\newcommand{\MAIF}{\mbox{MA$_{\rm{IF}}$}}

\newcommand{\finity}{\phi_{\rm{fin}}}

\newcommand{\IF}{image-finiteness}

\def\arr#1{\:\stackrel{#1}{\mbox{\rightarrowfill}}\:}

\newcommand{\Ar}[1]{\:\stackrel{{#1}}{\Longrightarrow} \:}

\newcommand{\Stat}[2]{{"={^{(#1 ,  #2)^\star}}!>"}}  
\newcommand{\StatOI}[1]{\Stat{\outamb{n}}{\inamb n}}
\newcommand{\StatIO}[1]{\Stat{\inamb n}{\outamb n}}

\def\reff#1{(\ref{#1})}       

\newcommand{\bisMOD}{\mathrel{\simeq_{\mathrm{int}}}} 

\renewcommand{\tilde}{\widetilde}

\mathcode`\!="4021       
\mathcode`\|="326A       
\mathcode`\.="602E       

\renewcommand{\true}{\ltrue}

\newcommand{\cn}[1]{\Ftame{#1}.}

\newcommand{\N}{\mathbb{N}}

\newcommand{\redin}[1]{\mbox{$\xrightarrow{\mbox{\inamb{#1}}}$}}

\newcommand{\F}{\mathcal{F}}
\newcommand{\R}{\mathcal{R}}

\newcommand{\non}{\lnot}

\newcommand{\Ftext}[1]{\ensuremath{\mathsf{#1}}}
\newcommand{\Fmeta}[1]{\may{#1}}
\newcommand{\Ftame}[1]{\must{#1}}

\newcommand{\A}{\AAA}

\input{commonmacro.sty}
\input{newmacro.sty}

\newcommand{\cal}[1]{\mathcal{#1}} 

\def\mths{\mathsurround=0 pt}
\def\eqalign#1{\null\,\vcenter{\openup\jot\mths
  \ialign{\strut\hfil$\displaystyle{##}$&$\displaystyle{{}##}$\hfil
      \crcr#1\crcr}}\,}
\def\Qed{\pushright{\qEd}
    \penalty-700 \par\addvspace{\medskipamount}}

\def\doi{2 (2:3) 2006}
\lmcsheading%
{\doi}
{35}
{}
{}
{Apr.~\phantom{0}4, 2005}
{Mar.~24, 2006}
{}

\begin{document}

\title[On the Expressiveness of the Ambient Logic]{On the Expressiveness of the Ambient Logic}

\thanks{Work
    supported by european project FET-Global computing
    PROFUNDIS.}
    
 \thanks{This work is a revised and extended version of parts
of~\cite{Sangiorgi::ExtInt::01} and~\cite{Sedal} (precisely, those
parts that deal with expressiveness issues).}

\author[D.~Hirschkoff]{Daniel Hirschkoff\rsuper a}
\address{{\lsuper a}LIP - ENS Lyon}
\email{Daniel.Hirschkoff@ens-lyon.fr}

\author[{\'E}.~Lozes]{{\'E}tienne Lozes\rsuper b}
\address{{\lsuper b}LSV- ENS Cachan}
\email{Etienne.Lozes@ens-lyon.fr}

\author[D.~Sangiorgi]{Davide Sangiorgi\rsuper c}
\address{{\lsuper c}Universit{\`a} di Bologna}
\email{Davide.Sangiorgi@cs.unibo.it}

\keywords{Spatial Logics, Mobile Ambients}

\begin{abstract} 
  The Ambient Logic (AL) has been proposed for expressing
  properties of process mobility in the calculus of Mobile Ambients
  (MA), and as a basis for query languages on semistructured data.
  
  In this paper, we study the expressiveness of AL. 
We define
  formulas for capabilities and for communication in MA. We also
  derive some formulas that capture finitess of a term, name
  occurrences and persistence. We study extensions of the calculus
  involving more complex forms of communications, and we define
  characteristic formulas for the equivalence induced by the logic on
  a subcalculus of MA. This subcalculus is defined by imposing an
  image-finiteness condition on the reducts of a MA process.
  \end{abstract}

\maketitle

\tableofcontents
\vskip-\bigskipamount

\section{Introduction}
The \emph{Ambient Logic}, AL, \cite{Cardelli::Gordon::AnytimeAnywhere}
is a modal logic for expressing properties of processes in the
calculus of Mobile Ambients, \MA\ \cite{CaGo98,CaGo99popl}.  In \MA\
the unit of movement is an ambient, which, intuitively, is a named
location.  An ambient may contain other ambients, and
\emph{capabilities}, which determine the ambient movements.  The
primitives for movement allow: an ambient to enter a sibling ambient;
an ambient to exit the parent ambient; a process to dissolve an
ambient boundary. \MA\ has a replication operator to make a process
persistent, that is, to make infinite copies of the process available.

An ambient can be thought of as a labelled tree. The sibling relation
on subtrees represents spatial contiguity; the subtree relation
represents spatial nesting.  A label may represent an ambient name or
a capability; moreover, a replication tag on labels indicates the
resources that are persistent.\footnote{We are using a tree
  representation different from that of Cardelli and Gordon, but 
more convenient to
  our purposes.}  The trees are unordered: the order of the children
of a node is not important.  As an example, the process $ P \defiDS !
\amb a {\inamb c} | \openamb a . \amb b \nil$ is represented by the
tree:\\[.1em]
\mbox{}\hfill$
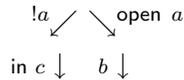
  \begin{array}[t]{ccc}
    &&\\
& {}^{!a}\!\!\swarrow\,\,  \searrow^{\openamb a} & \\[\tkpPOPL]
&\hskip -1cm  \mbox{{\scriptsize $\inamb c 
$}} \downarrow\hskip .3cm   \mbox{{\scriptsize $ b 
$}} \downarrow 
\end{array}$\hfill\mbox{}\\[.6em]
The  replication $!a$ indicates that the resource 
$ \amb a {\inamb c} $ is persistent: unboundedly many such ambients 
 can be spawned. By contrast, $\openamb a$ is
ephemeral: it can open only one ambient.

Syntactically, each tree is finite. Semantically, however, due to
replications, a tree is an infinite object.  As a consequence, the
temporal developments of a tree can be quite rich.  The process $P$
above (we freely switch between processes and their tree
representation) has only one reduction, to $\inamb c | ! \amb a
{\inamb c} | \amb b \nil$.  However, the process $! \amb a {\inamb c}
|! \openamb a . \amb b \nil$ can evolve into any process of the form
\maths{ \inamb c | \ldots | \inamb c | \amb b \nil | \ldots | \amb b
  \nil | ! \amb a {\inamb c} | ! \openamb a . \amb b \nil\,.}
In general, a tree may have an infinite temporal branching, that is,
it can evolve into an infinite number of trees, possibly quite
different from each other (for instance, pairwise behaviourally
unrelated).  Technically, this means that the trees are not
\emph{image-finite}.
 
 In summary, \MA\ is a calculus of dynamically-evolving unordered
 edge-labelled trees,  and    AL is a logic for reasoning on
 such trees.    The actual definition of satisfaction of the
 formulas of AL is given on \MA\ processes quotiented by a relation of
 \emph{structural congruence}, 
which equates processes with
 the same tree representation. (This relation is similar to
 Milner's structural congruence for the $\pi$-calculus~\cite{Mil99}.)
 
 AL has also been advocated as a foundation of query
 languages for semistructured
 data~\cite{CardelliSemiStructuredData}. 
 Here, the laws of the logic are used to describe query rewriting
 rules and query optimisations.  This line of work exploits the
 similarities between dynamically-evolving edge-labelled trees
 and standard models of semistructured data.

 AL has a   connective that talks about
\emph{time}, that is, how   processes  can evolve: the  formula 
$\diamond \AAA$ is satisfied by those processes 
with a   future in which  $\AAA$ holds. 
The logic has also connectives that talk about \emph{space}, that is,
the shape of the edge-labelled trees that describe process
distributions:
 the formula $\amb n \AAA$ is satisfied by  
ambients  named $n$ 
whose content  satisfies $\AAA$ (read on
trees: $\amb n \AAA$ is satisfied by the trees whose root has just a 
single edge
$n$ leading to a subtree that satisfies $\AAA$); 
 the formula $\AAA_1 | \AAA_2$ is satisfied by the processes 
that can be decomposed  into  parallel components 
$P_1$ and $P_2$ where each $P_i$ satisfies   $\AAA_i$ (read on
trees: $\AAA_1 | \AAA_2$ is satisfied by the trees that are the
juxtaposition of  
two trees that respectively satisfy the formulas $\AAA_1$ and $\AAA_2$);
the formula $\zero$ is satisfied by  the terminated   process $\nil$ (on
trees: $\zero $ is satisfied by the  tree consisting of  just the root node).

AL is quite different from standard modal logics.  First, such logics
do not talk about space. Secondly, they have more precise temporal
connectives.  The only temporal connective of AL talks about the
many-step evolution of a system on its own. In standard modal logics,
by contrast, the temporal connectives also talk about the potential
interactions between a process and its environment. For instance, in
the Hennessy-Milner logic~\cite{HeMi85}, the temporal modality
$\mess{\mu}.\AAA$ is satisfied by the processes that can perform the
action $\mu $ and become a process that satisfies $\AAA$.  The action
$\mu $ can be a reduction, but also an input or an output.
The lack of temporal connectives in
the ambient logic is particularly significant because in \MA\
interaction between a process and its
environment can take several forms,
originated  by
the  communication and the movement
primitives. (There are 9 such forms; they
appear as labels of transitions in a purely SOS semantics of \MA\
\cite{CaGo98annex,LeSa00full}.) 

\bigskip

This paper is essentially devoted to the study of the expressiveness
of AL.
The results we present show that AL is actually a very expressive
formalism. In particular, we are able to derive formulas expressing
capabilities of processes for movement and for communication, as well
as the persistence of processes (as given by the replication
operator), and free occurrences of names in processes.  The ability to
derive such constructions is surprising, considering that there
is no connective in the logic that is directly related to such
properties: no construct mentions the capabilities of the calculus,
nor does the logic include infinitary operators, or operators that
talk about resources with infinite multiplicity.

\smallskip

 Our results are established using nontrivial technical
developments, and the methods we exploit are of interest in their
own. 
  More precisely, the general approach to derive  expressiveness
formulas is to exploit adjunct connectives to introduce a form of
contextual reasoning, together with the temporal modality to make it
possible to observe the desired properties.
It can be noted that related constructions have been introduced in the
setting of Separation Logic~\cite{Rey02} in order to express weakest
preconditions for pointer manipulation instructions in an imperative
language.

\medskip

The expressive power of AL that we thus prove has several
consequences. The first consequence is that we are able to define
\emph{characteristic formulas} for image-finite Ambient processes,
i.e., formulas that capture the equivalence class of a process with
respect to the induced logical equivalence.  This is in contrast with
usual results in modal logics. Typically, the definition of
characteristic formulas exploits fixed-point operators, and the
characterised processes are
finite-state~\cite{GrafSifakis,SteffenIngolfsdottir}. As mentioned
above, AL has no fixed-point operator; moreover the \IF\ condition on
processes is weaker than finite-state.  (`Image-finite' expresses
finiteness on internal reductions, whereas `finite-state' also takes
into account computations containing visible actions such as input and
output actions.)

\medskip

Another major consequence of our results is to show that AL is an
\emph{intensional} logic. Informally, this holds because the logic allows one
to inspect the structure of processes, not only by separating
subcomponents of a process, but also by capturing its interaction
capabilities. 
More formally, intensionality of the Ambient Logic is expressed by
showing that the equivalence induced by the logic coincides with
structural congruence on processes. This result, that is established
using the constructions we have discussed above (and, in
particular, characteristic formulas), says that AL is a very fine
grained logic.


\paragraph{Structure of the paper.}

Section~\ref{s:back} introduces the calculus and the logic we study in
this paper. Sections~\ref{s:capacom} and~\ref{s:intensional} present
two main contributions in terms of expressiveness of AL: we define
some formulas capturing respectively some syntactical constructions of
the calculus (capabilities for movement and communication) and some
nontrivial properties of processes (finiteness, occurrences of free
names, and persistence). In Section~\ref{s:chara}, we exploit these
constructions to define characteristic formulas for logical
equivalence.  Intensional bisimilarity, which, for the purposes of the
present work, is a technical device that is needed to reason about
characteristic formulas, is presented in
Subsection~\ref{subsec:intbis}. The proofs of the main properties
enjoyed by intensional bisimilarity are not provided, and can be found
in a companion paper~\cite{Part2}.
Finally, in Section~\ref{s:ext}, we study extensions of the calculus
we work with, and show our results can be adapted to the corresponding
settings.

The results of this paper come from the two conference papers
\cite{Sangiorgi::ExtInt::01} and \cite{Sedal}: in \cite{Sangiorgi::ExtInt::01}, the author presented the encoding
of the modalities for capabilities and communications (Sections \ref{s:capacom} and \ref{s:ext}) and
the definition of intensional bisimilarity,
whereas the formulas capturing finiteness, name occurrence, and persistence (Section 
\ref{s:intensional}) and the characteristic formulas (Section \ref{s:chara}) come from \cite{Sedal}.
This paper focuses on the expressiveness results coming from these two conference papers, whereas a companion paper \cite{Part2} presents the separability results.

\paragraph{Developments.} 

By the time the writing of the present paper was completed, a few
works have appeared that make use of results or methods presented
here. We discuss them below.

The `contextual games' we have discussed above have been exploited in
several settings. Along the lines of the derivation of formulas
capturing Mobile Ambients capabilities, ~\cite{hls::fsttcs::2003}
extends and develops this line of research in the setting of a
sub-logic of AL, that is applied to reason about MA and $\pi$-calculus
processes.
Other interesting properties can be derived using this approach. An
example is quantifiers elimination~\cite{caires::lozes::elim}. Another
study~\cite{hirschkoff::extensionalspatial04} demonstrates that in
some sense, contextual games represent the logical counterpart of
`contextual testing' as in barbed equivalence~\cite{SW01}.

Our expressiveness results also allow us to bring to light
redundancies in spatial logics for concurrency. For example, an operator
to express occurrences of free names in processes is analysed in
related
works~\cite{Cardelli::Gordon::NameRestriction::01,hls::fsttcs::2003}.
In the setting of the present work, such an operator is encodable in
AL.

This kind of encodability results allow one to compare different
versions of spatial logics for concurrency, and are useful to assess
minimality properties of the logics.



\section{Background}\label{s:back}
This section collects the necessary background for this paper. It
includes the MA calculus~\cite{CaGo98} (semantic and syntax), and the
Ambient Logic~\cite{Cardelli::Gordon::AnytimeAnywhere}.

\subsection{Syntax of Mobile Ambients}

We recall here the syntax of MA~\cite{CaGo98} (we sometimes call this
calculus the \emph{Ambient calculus}).
We first consider  the calculus in which only names, not capabilities, can be
communicated; this allows us to work in an untyped calculus.  We
analyse extensions of the calculus in Section~\ref{s:ext}.  

As in~\cite{Cardelli::Gordon::AnytimeAnywhere,Car99data,CaGh00semi}, the  calculus has no restriction
operator for creating new names.  
The restriction-free calculus has a more direct correspondence with
edge-labelled trees and semistructured data.

Table~\ref{ta:syn} shows the syntax. Both the set of names and that of
variables are infinite. Letters $n,m,h $ range over names, $x,y,z$
over variables; $\eta$ ranges over names and variables.  The
expressions $\inamb \eta$, $\outamb \eta$, and $\openamb \eta$ are the
\emph{capabilities}, and are ranged over using \capa.  Messages and
abstractions are the \emph{input/output} (I/O) primitives. 
The metavariables $M,N$, for messages, will 
become usefull when considering  extensions of the language
 (see Section~\ref{s:ext}).
A \emph{closed} process has no free variables. We ignore syntactic
differences due to alpha conversion, and we write $P\sub n x$ for the
result of substituting $x$ with $n$ in $P$.
In the paper, all definitions and results are given only for closed
processes, unless otherwise stated.

Given an integer $n>0$, we will write $P_i, (1\leq i\leq n)$ for a
(finite) sequence of processes $P_1,\dots, P_n$.

\ifpopl
\begin{table*} 
\else
\begin{table} 
\fi
\begin{center}
\begin{tabular}{cc}
$
\begin{array}[t]{lrll}
\multicolumn{3}{l}{
  h,k, \ldots n,m}
& {\mbox{{\it Names}}} 
\\[\tkp]
  \multicolumn{3}{l}{
   \eta}
& {\mbox{{\it Names $\cup$ Variables}}} 
\\[\tkp]
\\[\tkp]
 &&& {\mbox{\it Expressions}}
\\[\tkp] 
M,N & ::=
&\capa & \mboxB{capability}
\\[\tkp]
\\[\tkp]
 &&& {\mbox{\it Capabilities}}
\\[\tkp] 
\capa & ::=
   &  \ina \eta
   & \mboxB{enter}
\\[\tkp]
 & \midd
   &  \out  \eta
   & \mboxB{exit}
\\[\tkp]
 & \midd
   &  \open  \eta
   & \mboxB{open}
\end{array}
$
\qquad& 
$\begin{array}[t]{lrll}
 &&& {\mbox{\it Processes}}
\\[\tkp] 
P,Q,R & ::=  
   & \nil
   &
   \mboxB{nil}
\\[\tkp]
 & \midd
   & P |  Q
   & \mboxB{parallel}
\\[\tkp]
 & \midd
   & !P
   & \mboxB{replication}

\\[\tkp]
 & \midd
   & M . P
   & \mboxB{prefixing}
\\[\tkp]
 & \midd
   & \amb{ \eta}{P}
   & \mboxB{ambient}
\\[\tkp]
 & \midd
   & \msg{ \eta}
   & \mboxB{message}
\\[\tkp]
 & \midd
   & \abs{x} P
   & \mboxB{abstraction}
\end{array} 
$
\end{tabular}
\end{center} 

 \caption{The syntax of finite MA}
\Mybar  
\label{ta:syn}
\ifpopl
\end{table*} 
\else
\end{table} 
\fi

Processes having the same internal structure are identified. This is
expressed by means of the \emph{structural congruence relation},
$\equiv$, the smallest congruence such that:

\begin{mathpar}

  P | \nil ~\equiv~ P
  \and
  P | Q ~\equiv~ Q | P
  \and
  P | (Q|R)~ \equiv~ (P | Q) | R
  \\
    !P ~ \equiv ~ !P | P
    \and
    !\nil ~ \equiv ~ \nil
    \and
    !(P|Q) ~ \equiv ~ !P | !Q
    \and
    !!P ~ \equiv ~ !P
\end{mathpar}

%

As a consequence of  results  in~\cite{Dal01}, that
studies a richer calculus than the one we study, we have:
\begin{thm}\label{thm:equivdeci}
\label{t:dz}
  The relation $\equiv$ is decidable.\Qed
\end{thm} 

The two following syntactic notions will be useful below.

\begin{definition}[Finite and single
  processes]\label{def:finitesingle}\mbox{}
  \begin{itemize}
  \item A closed process $P$ is \emph{finite} if there exists a
    process $P'$ with no occurrence of the replication operator such
    that $P\,\equiv\, P'$.
  \item A closed process $P$ is \emph{single} if there exists $P'$
    such that either $P\,\equiv\,\capa.P'$ for some \capa{}, or
    $P\,\equiv\,\amb{n}{P'}$ for some $n$, or $P\,\equiv\,(x)P$ for
    some $x$.
  \end{itemize}
%
\end{definition}

Unless otherwise stated, all results and definitions we state in the
sequel are on closed terms.


\subsection{Operational Semantics}

The operational semantics of the calculus is given by a reduction
relation $\longrightarrow$, defined by the rules presented in
Table~\ref{t:red_rules}.  The reflexive and transitive closure of
$\longrightarrow $ is written $\Longrightarrow $.

\begin{table}
\label{t:red_rules}
$$\begin{array}{c}
 \shortaxiomC{ \open n . P |
\amb n{ Q} \longrightarrow 
P |  Q
}
{Red-Open} 
\quad
  \shortaxiomC
{ \amb n {\ina m . P_1 | P_2} | 
\amb m{ Q} \longrightarrow 
\amb m{ \amb n { P_1 | P_2}| Q
}}{Red-In}
 \\[\tkpP]
 \shortaxiomC
{
\amb m{ \amb n {\out m . P_1 | P_2}| Q}
\longrightarrow 
 \amb n {P_1 | P_2} | 
\amb m{ Q} 
}
{Red-Out}  
\qquad\quad
\shortinfruleN
{
 P   \longrightarrow  P'
}
{
\amb n P   \longrightarrow  \amb n {P'}
}
{Red-Amb}
\\[\tkpP]
 \shortaxiomC
{ \msg M |
\abs x  P \longrightarrow 
P \sub M x
} 
{Red-Com}
 \quad\quad
\shortinfruleN
{
 P  \longrightarrow  P'
}
{
P |Q  \longrightarrow  P' |Q
}
{Red-Par}
\quad\qquad
 \\[\tkpP]
\shortinfruleN
{
 P \equiv P' \andalso P'  \longrightarrow  P'' \andalso P'' \equiv P'''
}
{
P   \longrightarrow  P'''
}
{Red-Str}
\end{array} 
$$
\caption{The rules for reduction}
\Mybar  
\end{table}

\begin{lem}
\label{lem:red-form}
If $P\longrightarrow Q$ then there is a derivation of the reduction in
which \trans{Red-Str} is applied, if at all, only as the last rule.\Qed
\end{lem}

Lemma~\ref{lem:red-form} shows that every reduction $P \longrightarrow 
P'$ has a normalised derivation proof.  As a consequence, we have:

\begin{lem}
\label{lem:basic-reduction}
If $P \longrightarrow  Q$ then either
\begin{enumerate}
\item
$P\equiv
R | 
\amb m{ \amb n {\out m . P_1 | P_2}| P_3}
$ and $Q \equiv
R|  \amb n {P_1 | P_2} | 
\amb m{P_3} 
$, or

\item
$P\equiv
R |
 \amb n {\ina m . P_1 | P_2} | 
\amb m{ P_3} 
$ and $Q \equiv
R| 
\amb m{ \amb n { P_1 | P_2}| P_3} 
$, or

  \item
$P\equiv
R |
 \open n . P_1 |
\amb n{ P_2}
 $ and $Q \equiv
R| 
P_1 | P_2
$, or

\item $P\equiv
R |
 \msg n |
\abs x { P_1} 
$
and $Q \equiv
R| 
P_1 \sub n x
$, or

\item $P\equiv
R | \amb n {P_1}$, $Q\equiv R |\amb n {Q_1}$ and $P_1 \longrightarrow Q_1$.\Qed

\end{enumerate}
\end{lem}


We now introduce some forms of labelled transitions that we will use
to give the interpretation of some of our logical constructions.

\begin{definition}[Labelled transitions]
\label{d:statt}
Let $P$ be a closed process. We write: 
\begin{itemize}
\item
$P \arr{\capa} P'$, where $\capa$ is a capability, if 
$P \equiv  \capa.P_1|P_2$ and $P' = P_1|P_2$.

\item $P \arr{\msg n} P'$ if 
$P \equiv  \msg n|P'$.

\item $P \arr{?n} P'$ if 
$P \equiv  \abs x {P_1}|P_2$ and $P' \equiv  P_1\sub n x|P_2$. 

\item $P \Ar{\mu}P'$, where $\mu$ is one of the above labels, if $P
  \Longrightarrow \arr\mu \Longrightarrow P'$ 
(where $
  \Longrightarrow \arr\mu \Longrightarrow$ is relation composition). 

\item {\bf (stuttering)}  $P  \Stat{\capa_1}{\capa_2} P' $ if 
there is $i\geq 1$ and processes $P_1 , \ldots, P_i$ with $P = P_1 $
and $P' = P_i$  such that 
$P_r \Ar {\capa_1}\Ar {\capa_2}P_{r+1}$ for all $1 \leq r < i$. 

\item Finally, $\Rcap$ is a convenient notation for compacting
  statements involving capability transitions. We let
  $\Rcapar{\inamb{n}}$ stand for $\Stat{\outamb{n}}{\inamb{n}}$;
  similarly $\Rcapar{\outamb{n}}$ is $\Stat{\inamb{n}}{\outamb{n}}$;
  and $\Rcapar{\openamb{n}}$ is $\Rar$.
\end{itemize}
\end{definition}

\subsection{The Ambient Logic}

\begin{table*}[t]
\begin{center}
$
\begin{array}[t]{lrlll}
\AAA & ::=
   &  \ltrue 
   & \mboxB{true}
   & 
\\[\tkp]
&\midd&  \neg\AAA
   & \mboxB{negation}
\\[\tkp]
&\midd&  \AAA \lor \BBB
   & \mboxB{disjunction}
\\[\tkp]
&\midd& \all x \AAA
   & \multicolumn{2}{l}{\mboxB{universal quantification over names}}
   \\[\tkp]
   & \midd &
   \diamond\AAA
&\mboxB{sometime}
&
\\[\tkp]
&\midd& \zero
&\mboxB{void}
\\[\tkp]
&\midd& \amb\eta\AAA
&\mboxB{edge}
\\[\tkp]
&\midd& \AAA | \BBB
&\mboxB{composition}
\\[\tkp]
& \midd &
\at{\AAA}{\eta}
&\mboxB{localisation}
&
\\[\tkp]
&\midd &\AAA~\rtr~\BBB
   & \mboxB{guarantee}\qquad
\end{array}
$
\end{center} 
\caption{The syntax of logical formulas}
\Mybar  
\label{ta:for}
\end{table*}

The logic has the propositional connectives, $\ltrue, \non \AAA, \AAA
\lor \BBB$, and universal quantification on names, $\forall x.\,
\AAA$, with the standard logical interpretation.  The temporal
connective, $\sometime \AAA$ has been discussed in the introduction.
The spatial connectives, $\zero$, $\AAA | \BBB$, and $\amb \eta \AAA$,
are the logical counterpart of the corresponding constructions on
processes. $\AAA \rtr \BBB $ and $\at{\AAA}{\eta}$ are the logical
adjuncts of $\AAA | \BBB$ and $ \amb \eta \AAA$ respectively, in the
sense of being roughly their `contextual inverse', as expressed in
Definition~\ref{d:satisfaction} below.

The logic in~\cite{Cardelli::Gordon::AnytimeAnywhere} has also a \emph{somewhere} connective,
that holds of a process containing, at some arbitrary level of nesting
of ambients, an ambient whose content satisfies $\AAA$.  We do not
consider this connective in the paper because we find it less
fundamental than the other operators; in any case, its addition would
not affect the results in the paper and has been seldomly considered
in other works. (Further, we discuss in the final section
a ``strong'' version of the sometimes modality.)

\begin{definition}[Satisfaction]
\label{d:satisfaction}
The satisfaction relation between closed processes and closed formulas,
written $\satDS P \AAA$, is defined as follows:
$$
\begin{array}{lcl}
\satDS P \true &\defiDS & \mbox{always true} 
\\
        \satDS P{\all x \AAA } & \defiDS & \mbox{for any $n$, } \satDS
        P \AAA \{ n /x\} 
        \\
\satDS P{ \myneg \AAA } & \defiDS & \mbox{not }  \satDS P{  \AAA }
\\
        \satDS P{ \AAA_1 | \AAA_2 } & \defiDS & 
                \exists P_1,P_2 \mbox{ s.t. } 
                P \equiv {P_1|P_2}
                \mbox{ and } \satDS{P_i}{\AAA_i},~i=1,2
        \\
        \satDS P{\AAA \orr \BBB } & \defiDS & \satDS P{\AAA } \mbox{ or }
        \satDS P{\BBB } 
\\
        \satDS P{\amb n \AAA } & \defiDS & 
                \exists P' \mbox{ s.t. }P \equiv \amb n {P'}
                \mbox{ and } \satDS{P'}\AAA
        \\
\satDS P{\zero } & \defiDS & P\equiv\nil
\\

        \satDS P{\sometime \AAA  } & \defiDS & 
                \exists P' \mbox{ s.t. }
                 P \Rar  P' \mbox{ and } \satDS{P'}\AAA 
        \\
        \satDS P{\at{\AAA}{n}   } & \defiDS & 
                \satDS{\amb n P } \AAA
\\
                \satDS P{\AAA \limp \BBB    } & \defiDS & 
                \forall R,~ R\sat\AAA \mbox{ implies } \satDS{P|R} \BBB
\end{array}
$$
\end{definition}

By definition, satisfaction is closed by structural congruence:

\begin{lem} 
  If $P \equiv Q $ and $\satDS P \AAA$, then also $\satDS Q \AAA$.\Qed
\end{lem}

We give $\lor $ and $\land$ the least syntactic precedence, thus
$\AAA_1 \rtr \AAA_2 \land \AAA_3$ reads $(\AAA_1 \rtr \AAA_2) \land
\AAA_3$, and $\AAA_1 \rtr (\sometime \AAA_2 \land \sometime \AAA_3)$
reads $\AAA_1 \rtr ((\sometime \AAA_2) \andd (\sometime \AAA_3))$.
We shall use the dual of some connectives, namely the duals of linear
implication ($\AAA \brtr \BBB$), of the sometime modality ($\square
\AAA$), of the parallel operator ($\|$), and the standard duals of
universal quantification ($\thereis x \AAA$) and disjunction ($\AAA
\land \BBB$); we also define (classical) implication ($\AAA\rightarrow 
{\cal B}$):

\begin{mathpar}
        \AAA \land B~\defiDS~  \non (\non \AAA \orr \non \BBB)
        \and
        \square \AAA~\defiDS~ \non \sometime \non \AAA 
        \and
        \AAA \to \BBB ~\defiDS~ \non \AAA \lor \BBB
        \and
        \thereis x \AAA ~\defiDS~ \non \all x \non \AAA 
        \and
        \AAA \brtr \BBB ~\defiDS~\non (\AAA \rtr \non \BBB)
        \and
                \lfalse ~\defiDS~ \non \ltrue
\end{mathpar}

Thus $\satDS P {\AAA \brtr \BBB}$ iff there exists $Q$ with $\satDS
Q\AAA$ and $\satDS {P|Q} { \BBB}$, and $\satDS P {\Box \AAA}$ iff
$\satDS{P'}{\AAA}$ for all $P'$ such that $P \Longrightarrow P'$.  

We now define the induced equivalence between processes induced by the
logic:

\begin{definition}[Logical equivalence]
\label{d:eqLOGIC}
For processes $P$ and $Q$, we write $P \eqL Q $ if for any closed
formula $\AAA$ it holds that $\satDS P \AAA $ iff $\satDS Q \AAA$.
\end{definition}



\section{Formulas for capabilities and
  communications}\label{s:capacom} 
In this section, we show that we can capture at a logical level
prefixes of the language, both for movement and for communication. 

\subsection{Preliminary formulas: counting components and comparing
names}

We start by recalling some formulas
from~\cite{Cardelli::Gordon::AnytimeAnywhere} that will be useful for
some constructions presented below.  

The Ambient Logic allows one to count the number of parallel
components of a process.  The formula below is true of a process that
has exactly one parallel component that is different from $\nil$.
\myframe{ \Ucomp \defi \myneg (\myneg \zero | \myneg \zero ) \land
  \myneg\zero }
\begin{lem}
\label{l:Ucomp}
It holds that
 $\satDS P \Ucomp$ iff  $P$ is single.\Qed 
\end{lem} 


Similarly we define 
\myframe{
 \Dcomp \defi \Ucomp | \Ucomp  
}

We may impose a given  formula \AAA{} to be satisfied by all single
parallel components of a process, using the following definitions:

\myframe{ 
  \begin{array}{rcl}
\AAA^\forall & \defi& \non(\non \AAA ~|~ \ltrue)
\\[\tkpPOPL]
\AAA^\omega& \defi& (\Ucomp \rightarrow \AAA)^\forall
  \end{array}
}

\begin{lem}
  \label{l:Aomega}\mbox{}
  \begin{itemize}
  \item $P\sat\AAA^\forall$ iff for any $Q,R$ such that
    $P\,\equiv\,Q|R$, it holds that $Q\sat\AAA$.
\item $P\sat\AAA^\omega$ iff all single parallel components of $P$
  satisfy \AAA.\Qed
  \end{itemize}
\end{lem}


We shall use later the following derived formula,
from~\cite{Cardelli::Gordon::AnytimeAnywhere}, that expresses equality
between names:

\myframe{ m=n\defi (\amb{n}{\ltrue})@m}

\begin{lem}
  $P\sat\,m=n$ iff names $m$ and $n$ are equal.\Qed
\end{lem}

\subsection{Formulas for capabilities}

The two formulas below are true of a process that is (structurally
congruent to) an ambient and (to)
an empty ambient, respectively.

\myframe{ 
\begin{array}{rcl}
\Uamb &\defi& \thereis x \amb x \ltrue  \\[\tkpPOPL]
\UambZero&\defi& \thereis x \amb x \zero 
\end{array}
}

\begin{lem}
\label{l:Uamb}\mbox{}
\begin{itemize}
\item
$\satDS P \Uamb$ iff $P \equiv \amb n Q$, for some $n $ and $Q$.

\item 
$\satDS P \UambZero$ iff $P \equiv \amb n \nil$, for some $n $.\Qed
\end{itemize}
\end{lem} 

To help understanding   the definitions of the capability
formulas, we  first discuss some   simpler  formulas, which 
 do not talk about
 the process  underneath the prefix. 
We define, for names $n \neq h$:
\[
\begin{array}{rcl}
\UopenN n &\defi&
\begin{array}[t]{l}
\amb n {\amb h  \zero} \limp  \diamond (\amb h \zero | \true) \\[\tkpPOPL]
\andd \; \Ucomp \\[\tkpPOPL]
\andd \; \myneg \Uamb
\end{array} \\[\tkpPOPL]
\UoutN n &\defi&
\begin{array}[t]{l}
\at{\at{ (\diamond (\amb h {\true} | \amb n \nil) )} n} h \\[\tkpPOPL]
\andd \; \Ucomp \\[\tkpPOPL]
\andd \; \myneg \Uamb
\end{array}
\end{array}
  \]

It holds that 
 $\satDS P {\UopenN n}$ iff $P \equiv \open n .P'$ for some $P'$. 
We sketch the proof.
The sub-formula  $\Ucomp 
\,\andd \; \myneg \Uamb$ says that $P$ is  single and is
not an ambient. Thus, modulo $\equiv$, process  $P$ can only 
be  $\nil$, $\open m . P'$, $ \ina m. P'$, $\out m . P'$, $\abs x
{P'}$, 
or $ \msg m$, for some $m$. 
The sub-formula  $\amb n {\amb h  \zero} \limp  \diamond (\amb h \zero
| \true) $ says that $P | \amb n {\amb h  \nil}$ can reduce to a
process with an empty ambient $h$ at the outermost level.  From 
 these  requirements, we conclude that $P \equiv
\open n .P'$, for some  $P'$.

Similarly we prove that
$\satDS P \UoutN n $ iff $P \equiv \out n . P'$, for some $P'$.
By the sub-formula $  \Ucomp
\: \andd \; \myneg \Uamb$, process $P$ is  single and is not an
ambient. 
By the sub-formula 
$\at{\at{ (\diamond (\amb h {\true} | \amb n \nil) )} n} h $,
\[\satDS{\amb {n}{\amb h P}}{\diamond (\amb h \true | \amb n \zero)} \] 
hence 
 $P \equiv\out n . P'$, for some $P'$, otherwise $\amb h P$ could not 
 exit  $n$.

To obtain  the full capability formulas
we add some quantification on names. Formula $\UopenN n . \AAA $ is
thus defined as follows:

\myframe{
\begin{array}{rcl}
\UopenN n . \AAA &\defi&
\begin{array}[t]{l}
\all y{ \amb n {\amb y  \zero} \limp  \diamond (\amb y \zero | \AAA)} \\[\tkpPOPL]
\andd \; \Ucomp \\[\tkpPOPL]
\andd \; \myneg \Uamb
\end{array}
\\[\tkpPOPL]
\Uopen &\defi& 
\begin{array}[t]{l}
\thereis x \all y{ \amb x {\amb y  \zero} \limp  \diamond (\amb y
  \zero | \true)} \\[\tkpPOPL] 
\andd \; \Ucomp \\[\tkpPOPL]
\andd \; \myneg \Uamb
\end{array}
\end{array}
}

\begin{remark}[Formulas containing free variables]
  It will often be the case in the remainder of the paper that we
  define a formula involving a name, say $n$, and need the
  corresponding logical construction where a variable $x$ is used
  instead of $n$. For instance, the formula $\Uopen$ above could be
  defined as as `` $\thereis x {\UopenN x} . \true$'', which is not
  correct because $\UopenN n.\AAA$ has been defined but $\UopenN
  x.\AAA$ has not.
  In the sequel, when clear from the context, we shall allow ourselves
  to adopt nevertheless this abuse of notation, that should be
  understood as `rewrite the definition of the corresponding formula
  using $x$ instead of $n$' (see in particular the formulas to capture
  name reception, and their interpretation, in
  Lemma~\ref{l:inputformulas}, and characteristic formulas for input
  guarded processes in Section~\ref{s:chara}).
  
  Satisfaction being defined only between closed processes and closed
  formulas, the important point in doing so is to avoid reasoning
  about the satisfaction of formulas containing free variables: we
  shall therefore only write formulas containing an `$x$' under the
  scope of a variable quantification.
\end{remark}

\begin{lem}
\label{l:Uopen}
$\satDS P \UopenN n . \AAA $ iff $P \equiv \open n . P'$, for some $P'$
such that $P' \Longrightarrow P''$ and $\satDS {P''}A$.

$\satDS P \Uopen$ iff $P\equiv \open n.P'$ for some $n$ and $P'$.
\end{lem} 

\proof
  We only consider the first property, from which the second follows
  easily. The implication from right to left is easy.

For the reverse implication, we set
\[G \defi \amb n {\amb h  \zero} \limp  \diamond (\amb h \zero |
\AAA) \]
where $h\not \in \n P$. 
Since $\satDS P \Ucomp$, we have  $P \equiv Q$, for some $Q$ that is not a
parallel composition.  Since also $\satDS P \myneg \Uamb$, we infer that
$Q$ is not an ambient.  Finally  since $\satDS P G$, process $Q$ cannot
be of the form $\nil, \ina n. Q', \out n . Q', \abs x {Q'}, \msg \MM$. 
For the same reason, $Q$ cannot be a prefix $\open m . Q' $ with  
 $m \neq n$.  The only possibility left is $Q = \open n . Q' $, for
 some $Q'$.

Moreover, we have
\[
\amb n {\amb h \nil} | \open n.  Q' \Longrightarrow R \mbox{ and } \satDS
R {\amb h \nil | \AAA  } 
 \]
for some $R$. 
The first step of this reduction must be 
\[ 
\amb n {\amb h \nil} | \open n . Q' \longrightarrow  {\amb h \nil} |  Q' 
\]
(up to $\equiv$). Since $h$ is fresh, ${\amb h \nil}$ cannot interact
with $Q'$. Hence 
\[ R \equiv {\amb h \nil} |  Q'' \]
for some $Q''$  such that $Q' \Longrightarrow Q''$.\Qed

Along the lines of our construction for the $\openname{}$ prefix, we
can define characteristic formulas for the $\inname{}$ and
$\outname{}$ prefixes.

\myframe{
\begin{array}{rcl}
\UoutN n . \AAA &\defi&
\begin{array}[t]{l}
\all x  \Big( \at{\at{ (\diamond (\amb x {\AAA} | \amb n \nil) )} n} x
\Big) \\[\tkpPOPL]
\andd \; \Ucomp \\[\tkpPOPL]
\andd \; \myneg \Uamb
\end{array}
\\[\tkpPOPL] 
\Uout &\defi& \thereis x {\UoutN x} \\
\ \\
\UinN n . \AAA &\defi&
\begin{array}[t]{l}
\all x \Big(
\at{ (\amb n \zero \limp \diamond \amb n {\amb x \AAA})} x\Big) \\[\tkpPOPL]
\andd \; \Ucomp \\[\tkpPOPL]
\andd \; \myneg \Uamb
\end{array}
\\[\tkpPOPL] 
\Uin &\defi& \thereis x {\UinN x}
\end{array}
}

\begin{lem}
\label{l:Uout}
$\satDS P \UoutN n . \AAA $ iff $P \equiv \out n . P'$, for some $P'$ such
that $P' \StatIO n P'' $ and $ \satDS{P''} \AAA$.
\end{lem} 

\proof
Similar to the proof for the \openname{} prefix.
The formula $  \; \Ucomp
\: \andd \; \myneg \Uamb$ forces $P$ to be single and not an
ambient. Therefore $P \equiv Q$, for some $Q$ whose outermost operator
is not a parallel composition or an ambient. Then we should have
\[\satDS{\amb {n}{\amb h Q}}{\diamond (\amb h A | \amb n \zero)} \] 
This can only happen if $Q$ is of the form  $\out n . Q'$, for some
$Q'$ such that 
 $Q'  \StatIO n Q'' $ and $\satDS{Q''}{\AAA}$.\Qed

\begin{lem}
\label{l:Uint}
$\satDS P \UinN n . \AAA $ iff $P \equiv \ina n . P'$, for some $P'$ such
that $P' \StatOI n P'' $ and $ \satDS{P''} \AAA$.
\end{lem} 

\proof
Similar to the previous proofs.
The formula $  \; \Ucomp
\: \andd \; \myneg \Uamb$ forces $P$ to be single and not an
ambient. Therefore $P \equiv Q$, for some $Q$ whose outermost operator
is not a parallel composition or an ambient. Then we should have
\[\satDS{\amb h Q| \amb n \nil}{\diamond  \amb n {\amb h \AAA}} \] 
where $h$ is fresh.
 As by previous arguments, 
this can only happen if $Q$ is of the form  $\ina n . Q'$, and $Q'$ reduces (with suttering) to $Q''$
satisfying $\AAA$.\Qed


Given a capability $\capa$, we may define the \emph{`necessity'} version
of the \emph{`possibility'} formulas we have just introduced as follows:

\myframe{
\Ftame{\capa}.\AAA \defi \Fmeta{\capa}.\ltrue~\land~
\non\Fmeta{\capa}.\non\AAA
}

\begin{lem}\label{capacformul} For any capability $\capa$, formula
  $\AAA$ and term $P$,
 $P~\sat~\Ftame{\capa}.\AAA$ iff there is $P'$ such that
    $P\,\equiv\,\capa.P'$, and, for any $P''$ such that
    $P'\,\Rcap\,P''$, $P''\sat\AAA$.
\end{lem}

Note that necessity formulas are not the dual of the possibility
formulas, as in standard modal logics, because of the spatial aspects
of AL. For instance, $ \cn {\inamb n} \true $ does not have the same
interpretation as $\neg \Fmeta{\inamb n}.  \neg\true$, the latter
being actually equivalent to $\ltrue$.

\begin{remark}
  We could think of deriving formulas for \emph{modalities} 
  $\Ar{\capa}\!$, as in standard modal logics for
  concurrency~\cite{HeMi85}, instead of capturing the syntactical
  prefixes corresponding to a capability $\capa$. More precisely, we could
  look for a formula $\Fmeta \capa\AAA$ 
  capturing processes $P$ for which there is $P'$ such that $P\Ar{\capa} P'$ and $P'\sat\AAA$. It turns out that
  spatial logics are more intensional, and make actions more difficult
  to express than connectives. In particular, we do not know how to
  express directly a modality corresponding to action $\Ar{\openamb
    n}$.
\end{remark}

\subsection{Formulas for communication}\label{subsec:comm}

The first step to characterise I/O processes (i.e., messages or
abstractions) is to get rid of other possible constructs for single
terms, as follows:

\myframe{
\begin{array}{rcl}
  \Ucomm &\defi& \Ucomp 
  \: \andd \myneg \Uamb  \: \andd \myneg \Uopen
  \: \andd \myneg \Uout \: \andd \myneg \Uin
\end{array}
}

\begin{lem}
\label{l:Ucomm}
$\satDS P \Ucomm$ iff 
($P \equiv \msg \MM$ or $P \equiv \abs x {P'}$), for some $\MM$
and $P'$.
\end{lem} 

The following formula, that holds of a process that is the parallel
composition of two I/O processes, will also be useful:

$$
  \begin{array}{rcl}
    \Dcomm &\defi&  \Ucomm | \Ucomm\,.
  \end{array}
$$

The difficult part, however, is the definition of the I/O formulas for
separating messages from abstractions, and also, within the messages
and the abstractions, messages with different contents and
abstractions with different behaviours.  

The capability formulas are easier to define than the I/O formulas
because capabilities act on ambients, and the logic has a connective,
$\amb n A$, for talking about ambients. By contrast, the I/O
primitives act on themselves.
To define the I/O formulas, we proceed as follows:
\begin{enumerate}
\item We define a formula,  $\TestComm$, that characterises the
  special  abstraction $\abs x{\amb x \nil}$.

\item We use  $\TestComm $ to define the
  formula 
for messages:
\[\UmsgN n \defi \Ucomm \: \andd (\TestComm \limp \diamond \amb n \zero) \]
\iflong
It holds that $\satDS P {\UmsgN n}$ iff $P \equiv \msg n$.
\fi
\item  We then use  $\UmsgN n$ to define the formulas 
  for abstractions:
\[
\UabsN n . \AAA \defi  \Ucomm \: \andd (\myneg \thereis x\UmsgN x )\: \andd 
(\UmsgN n \limp  \diamond \AAA  )
\]
\iflong
It holds that $\satDS P {\UabsN n . \AAA}$ iff $P \equiv \abs x Q $ and
$\msg n | P \Longrightarrow P'$ with $\satDS{P'}{\AAA}$.
\fi
\end{enumerate}

\begin{lem}
\label{l:mauam}
Given $\abs x R$, suppose  there is $\NN$
such that
\[ 
\satDS {\msg \NN  | \abs x R } {\Box (\Dcomm \orr \UambZero)}
 \]
 and $R$
contains no abstractions. Then $R \equiv \amb \eta \zero$, for some $\eta$.

\end{lem}

\medskip

We call \emph{ambient abstraction} any closed  abstraction described by the
following grammar:
\[ P ::= \abs x {\amb \eta \nil} \midd \abs x {(\msg \eta | P)}\] 
The following lemma shows how to characterise ambient abstractions
using formulas.

\begin{lem}
\label{l:auam}
Given an abstraction $\abs x R$,  
suppose  there is $\NN$ such that
\begin{equation}
\label{e:J1}
\satDS {\msg \NN  | \abs x { R }} {\Box (\Dcomm \orr \UambZero)}
\end{equation} 
and 
\begin{equation}
\label{e:J2}
\satDS {\msg \NN  | \abs x { R}} {\diamond  \UambZero}. 
\end{equation}
 Then $\abs x R$ is an ambient abstraction.
\end{lem}

\proof
By  induction on the number of nested abstractions in $R$.
 If this number is $0$ then  by 
 Lemma~\ref{l:mauam} we derive $R \equiv \amb \eta \zero$. 

Suppose the number is greater than $0$.  From \reff{e:J1} and 
\[\msg \NN  | \abs x { R } \longrightarrow R
\sub \NN x  \]
we derive 
\[ \satDS{R\sub \NN x }{
\Dcomm \orr \UambZero
} \] 
Since $R$ should contain an abstraction, the formula $\UambZero$ is
not satisfied, hence 
\[ \satDS{R\sub \NN x }{
\Dcomm 
} \]  
Using this, the fact that $R\sub \NN x$ should contain an abstraction, and
\reff{e:J2} we infer that 
\[ R\sub \NN x \equiv \msg{\MM} | \abs y Q\]
for some $\MM,y,Q$. By induction hypothesis, we deduce that $(y)Q$ is an ambient abstraction.
From this, $R\sub \NN x$ is an ambient abstraction too, and 
this induces that $R$ itself is an ambient abstraction. \Qed

We say that an ambient abstraction $P$ is \emph{simple} if $P =_\beta
\abs x {\amb x \zero}$,
 where $=_\beta$ is the least congruence that is
 closed under the rule
 \[ \msg M |  \abs x P  = P \sub M x\,. \]
 
 We recall that the operator $\limpE$, used in the following lemma,
 has been introduced at the end of Section~\ref{s:back}.

\begin{lem}
\label{l:auamSIMPLE}
Suppose  $\abs x Q$ is an ambient abstraction, and 
\[
\begin{array}{l}
\satDS{
 \abs x  Q }{\Ucomm\, \limpE\, \diamond \amb n \zero} \\[\tkpPOPL]
\satDS{
\abs x  Q }{\Ucomm\, \limpE\, \diamond \amb m \zero} 
\end{array} 
\]
with $m \neq n$. Then $\abs x Q$ is simple.
\end{lem} 

\proof From the hypothesis, there are $ \MM$ and $\NN$ such that 
\[
\begin{array}{l}
\satDS{\msg \MM |
 \abs x  Q }{ \diamond \amb n \zero}\, \quad\mbox{and} \\[\tkpPOPL]
\satDS{\msg \NN | 
\abs x  Q }{ \diamond \amb m \zero}\,. 
\end{array} 
\]
If $\abs x Q$ were not simple, then the name of the ambient to which 
it reduces to would not depend on the argument $x$.
(Note that any ambient abstraction is $=_\beta$ to an abstraction of
the form $\abs x {\amb \eta \zero}$, for some $x,\eta$. The  hypothesis of
the lemma implies that $\eta = x$.)\Qed

As hinted above, the key step is the definition of the formula below,
which is the characteristic formula of simple ambient abstractions.

\vskip .2cm 
\noindent\framebox{\parbox{426 pt}{
\begin{minipage}{426 pt}
\begin{eqnarray}
\nonumber 
\TestComm & \defi & 
\Ucomm  \\[\tkpPOPL]
&&
\label{eqn:F1} 
\andd \; \Ucomm \limp \Box (\Dcomm \orr \UambZero) \\[\tkpPOPL]
&&
\label{eqn:F2} 
\andd \; \Ucomm \,\limpE\, \diamond \amb n \zero \\[\tkpPOPL]&&
\label{eqn:F3} 
\andd \; \Ucomm \,\limpE\, \diamond \amb m \zero 
\end{eqnarray}
where $n,m$ are different names.
\end{minipage}
}}
\vskip .2cm 

\begin{lem}
\label{l:TestComm}
$\satDS P \TestComm$ iff $P$ is a simple ambient abstraction and is
closed.
\end{lem}  

\proof
The implication from right to left is easy. We consider the opposite.

Process $P$ must be an I/O, since $\satDS P \Ucomm$. Also, $P$
cannot be a message, otherwise it would not satisfy the formula 
\[  \Ucomm \limp \Box (\Dcomm \orr \UambZero)\,\]
\noindent since a message in parallel with $\abs x \nil$ can reduce to
$\nil$, which does not satisfy $\Dcomm \orr \UambZero$.

We conclude that $P$ should be an abstraction, say $\abs x Q$.  Now, from 
\reff{eqn:F1} and \reff{eqn:F2}, we get that there are messages  $ \MM,\NN$
such that 
\[
\begin{array}{l}
\satDS{\msg \MM | \abs x Q}{
\Box (\Dcomm \orr \UambZero)} \\[\tkpPOPL]
\satDS{\msg \NN | \abs x Q}{
 \diamond \amb n \zero }
\end{array} 
\]
From Lemma~\ref{l:auam} we infer that $\abs x Q$ is an ambient
abstraction. 
Moreover, by 
\reff{eqn:F2}, \reff{eqn:F3} 
and Lemma~\ref{l:auamSIMPLE}, $\abs x Q$   must be  simple.\Qed

Now we are finally in the position of defining the characteristic
formula for a message $\msg n$:
\myframe{
\UmsgN n \defi 
\begin{array}[t]{l}
\TestComm \limp \diamond \amb n \zero   \\[\tkpPOPL]
\andd\, \Ucomm
\end{array} 
}
and, then, the characteristic formula for a message is
\myframe{
  \mathtt{1mess} \defi
\thereis x \mathcal{F}_{\msg  x}
}

\begin{lem}\label{lem:charamsg}
\label{l:UcommNN}
$\satDS P{\mathcal{F}_{\msg n}}$ iff $P \equiv \msg n$, and $\satDS P
\mathtt{1mess}$ iff $P\equiv \msg n$ for some $n$.
\end{lem} 

\proof
  The right to left direction is easy. For the converse, we observe
  that $P$ must be an I/O, and that $P$ cannot be an abstraction
  (otherwise, when adding a process satisfying $\TestComm$, we could
  not obtain an ambient).  Hence $P \equiv \msg m$, for some $m$.  
  
  Given a simple ambient abstraction $Q$, we have that
$$ \satDS {Q | \msg m }{\diamond \amb n \nil} \quad
\mbox{iff}\quad m = n\,.
$$
This allows us to deduce that $P\equiv\msg n$.\Qed

We can now define the two modalities for the input connective:

\myframe{
\begin{array}{rcl}
\may{?n}.\A & \eqdef & \Ucomm \: \andd (\myneg \thereis x\UmsgN x )\: \andd 
(\UmsgN n \limp  \diamond \AAA  )
\\
\must{?n}.\A  & \eqdef & \may{?n}.\ltrue~\land~\non\may{?n}.\non\A
\\
\mathtt{1input} &\eqdef &\exists x.~\may {?x}.\ltrue
\end{array}
}

\begin{lem}\label{l:inputformulas}\mbox{}
  \begin{itemize}
  \item $P\sat \may{?n}.\A$ iff there are $P',P''$ such that $P\equiv
    (x)P'$, $(x)P'|\msg n\Rar P''$, and $P''\sat \A$.
  \item $P\sat \must{?n}.\A$ iff there is $P'$ with $P\equiv (x)P'$, and
    for all $P''$ such that $(x)P'|\msg n\Rar P''$, $P''\sat \A$.
  \end{itemize}
\end{lem}



\section{Other intensional properties}\label{s:intensional}
As we have just seen, AL can capture several
syntactical constructions of the calculus.   We now further explore the
expressiveness of AL,  going beyond the results  we have established about
capabilities and communications. 

We first define a formula $\phi_{fin}$ that characterises finite
terms, using a form of \emph{contextual reasoning}. The same method is
applied to derive a formula $\copyright n$ that characterises the
terms containing $n$ as a free name. We then introduce formulas that
characterise in a restricted sense \emph{persistent} single terms of
the calculus. These formulas will be used in Section~\ref{s:chara} to
establish characteristic formulas for a sub-calculus of MA.

\subsection{Capturing finiteness}

We now present a formula that is satisfied by all and only the finite
processes. Detecting replication seems \textit{a priori} unfeasible in
the present version of AL, as it does not provide a
recursion operator. We   capture the `finite' character of
a term using the fact that a replicated process is \emph{persistent},
i.e., it is always present along the reductions of a term.

The characterisation of finiteness relies on the existence of a
scenario which guarantees reachability of $\nil$, as expressed by the
two following lemmas:
\begin{lem}\label{finiteredcons}
  Let $P, Q$ be two terms such that $P\,\Rar\, Q$. Then $P$ is finite
  iff $Q$ is finite.
\end{lem}
\proof
  By induction over the length of the $\Rar$ derivation, then
  induction over the structure of the proof of the $\rar$ transition.\Qed

 \begin{lem}\label{bangintenslemma}
$ P$ 
is finite iff there are $Q, R, n$ such that
$\amb{n}{P\,|\,Q}~|~R~\Rar~ \nil$.
 \end{lem}

\proof\mbox{}

    \noindent$\bullet$ {Let us first assume that $P$ is finite}. We prove by
    induction on the size of $P$ that there exist $Q$ and $R$ such that for any
    $P'$,
    $$
    \amb{n}{P\,|\,P'\,|\,Q}~|~R~~\Rar~ \amb{n}{P'}
    $$
    The left to right implication can then be obtained using this
    property with $P'=\nil$ and adding $\openamb{n}$ in parallel with
    $R$.
    \begin{itemize}
    \item[1] For $P=\nil$, take $Q=R=\nil$.
    \item[2] For $P\equiv\amb{m}{P_1}$, we have by induction
      $Q_1,R_1$ such that $\amb{n}{P_1\Vert P'\Vert Q_1}\Vert R_1~\Rar
      \amb{n}{P'}$ for any $P'$.  Now we set $Q=\openamb{m}\,|\,Q_1$
      and $R=R_1$. Then it is clear that
      $\amb{n}{\amb{m}{P_1}\,|\,P'\,|\,Q}\,|\,R~\Rar~\amb{n}{P'}$ for
      any $P'$.
    \item[3] For $P~\equiv~P_1\,|\dots|\,P_r$ (with no
      replicated component), we use the induction hypothesis to obtain
      $Q_i$ and $R_i$, and then set $Q=Q_1\,|\dots|\,Q_r$,
      $R=R_1\,|\dots|\,R_r$ such that for any $P'$,
      $$
      \begin{array}{cl}
        \amb{n}{P\,|\,P'\,|\,Q}\,|\,R & \Rar~~
        \amb{n}{P_2\,|\dots|\,P_r\,|\,P'\,|\,Q_2\,|\dots|\,Q_r}
        \,|\,R_2\,|\dots|\,R_r \\
        & \Rar\dots\Rar~~\amb{n}{P'}
      \end{array}
      $$
      reasoning inductively on $r$.
    \item[4] For $P~\equiv~\capa.P_1$, we use the induction
      hypothesis to get $Q_1$ and $R_1$, and we define $Q$ and $R$
      according to the shape of $\capa$ as follows:
      \begin{itemize}
      \item $\capa=\inamb{m}$. Then we set $Q=Q_1$ and
        $R=\amb{m}{\nil}\,|\,\openamb{m}\,|\,R_1$. Then for any $P'$:
        $$
        \begin{array}{cl}
        \amb{n}{P\,|\,P'\,|\,Q}\,|\,R & \rar~
        \amb{m}{\amb{n}{P_1\,|\,P'\,|\,Q}}\,|\,\openamb{m}\,|\,R_1\\
        & \rar~\amb{n}{P_1\,|\,P'\,|\,Q_1}\,|\,R_1 \\
        & \Rar~\amb{n}{P'}
        \end{array}
        $$
      \item $\capa=\outamb{m}$.   We set $Q=\inamb{m}\,|\,Q_1$ and
        $R=\amb{m}{\nil}\,|\,\openamb{m}\,|\,R_1$, so that we can
        conclude.
      \item $\capa=\openamb{m}$. We set $Q=\amb{m}{\nil}\,|\,Q_1$
        and $R=R_1$.
      \end{itemize}
    \item For $P~\equiv~\msg m$, we set $Q=(x)\nil$, and $R=\nil$.
      \item For $P~\equiv~(x) P_1$: by induction hypothesis applied to 
        $P_1\{n/ x\}$, we get $Q_1$ and $R_1$; then we set $Q=\msg n|Q_1$
        and $R=R_1$.
    \end{itemize}
    The first implication is thus established.

  \noindent$\bullet$ {Let us now assume $P$ is not finite.}  Then for any $n,Q,R$,
    $\amb{n}{P|Q}|R$ is also infinite, and by the previous lemma, it
    is also the case for any of its reducts, and hence it cannot
    reduce to $\nil$.\Qed

We can now define:

\myframe{
\finity \defi \exists
    x.\big(\,\ltrue~\brtr~(\ltrue\,\brtr\,\sometime\zero)@x\,\big)
}

\begin{thm}
  For any $P$, 
    $P~\sat~\finity$ iff $P$ is  finite.
\end{thm}
\proof
From Definition~\ref{d:satisfaction}, $P\sat\finity$ holds if
 there are $n,Q,R$ such that
$n[P|Q]|R\Rar\nil$. We then conclude with Lemma~\ref{bangintenslemma}.\Qed

\subsection{Formula for name occurrence}\label{subsec:copyright}

Our aim is now to define a formula corresponding to the connective
$\copyright n$, defined by:
$$
P~\sat~\copyright n~~~\mbox{iff}~~~n\in\fn{P}\,.
$$

For this, 
 we exploit Lemma~\ref{l:freenames} together with the ability,
using the formulas for capabilities, to detect unguarded occurrences
of names.

We say that a process $P$ is \emph{flat} if it has no inputs and the only process
underneath all capabilities, and inside all ambients of $P$ is $\nil$. We
say that a process $P$ has an occurrence of name $n$ at top level if
$P\,\equiv\, \capa.P'|P''$ with $\capa = \inamb{n}$, $\outamb{n}$ or
$\openamb{n}$, $P\,\equiv\,\amb{n}{P'}|P''$ or $P\,\equiv\,\msg{n}|P'$.

For the proof of the next lemma, we would also need a more general
notion. The \emph{occurrence depth} of a name $n$ in an open term is
given by a function $\deep_n:\mathcal{P}\rar\N\cup\{\infty\}$, stable
by $\equivE$, inductively defined as follows:
\begin{itemize}
\item[-] $\deep_n(\nil)=\infty$.
\item[-] $\deep_n(\amb{n}{P_1})=0$, and for $n\neq \eta$,
  $\deep_n(\amb{\eta}{P_1})=\deep_n{P_1}+1$.
\item[-] $\deep_n((!)P_1\,|\dots|\,(!)P_r)=\min_{1\leq i\leq r}
  \deep_n(P_i)$ (here $(!)Q$ stands for $Q$ or $!Q$).
\item[-] $\deep_n(\capa.P)=
  \begin{cases}0&\text{for $\capa\in\{\inamb{n},\outamb{n},\openamb{n}\}$,}\\
    \deep_n(P)+1&\text{otherwise.}\end{cases}$
\item[-] $\deep_n((x)P)=\deep_n(\downarrow_\beta P)+1$, where $\downarrow_\beta P$ stands for the smallest term such that $P=_\beta\downarrow_\beta P$
\item[-]  $\deep_n(\msg n)=0$ and 
$\deep_n(\msg \eta)=\infty$ for $\eta\neq n$.
\end{itemize}

\begin{lem}
\label{l:freenames}
For all $P,n$, we have $n\in\fn{P}$ iff for any name $m$, there exist
some flat processes $Q,R$, in which $n$ does not occur free, and a
process $S$ with an occurrence of $n$ at top level such that
$ 
{
\amb{m}{P\,|\,Q}\,|\,R~\Rar~\amb{m}{S}
}
$.%
\end{lem}
\proof

Note that the property of $S$ having an occurrence of $n$ at top level
is equivalent to $\deep_n(S)=0$. We are now ready to prove the lemma:
\begin{itemize}
\item We first consider the implication from left to right.  Let us
  assume that $\deep_n(P)$ is finite.  We consider a name $m$, and prove
  by induction on $\deep_n(P)$ that there exist $Q,R,S$ satisfying the
  conditions of the lemma.
  \begin{itemize}
  \item if $\deep_n(P)=0$, we take $Q=R=\nil$ and $S=P$.
  \item if $\deep_n(P)=i+1$, we first consider the case where
    $P\equiv\inamb{m_1}.P_1\,|\,P_2$ with $\deep_n(P_1)=i$.  By
    induction hypothesis, there are $Q_1,R_1,S_1$ and $m$ satisfying
    the conditions of the lemma for $P_1|P_2$. 
    We then can set for
    $P$: $Q=Q_1$, $R=\amb{m_1}{\nil}\,|\,\openamb{m_1}\,|\,R_1$ and
    $S=S_1\,|\,P_2$, then $Q,R,S$ can be chosen for $P$.

    The other cases are treated similarly: we define processes that
    allow us to trigger a capability in order to decrease the
    occurrence depth of $n$ in the term. The definition of these
    processes follows the ideas in the proof of
    Lemma~\ref{bangintenslemma}.
    \end{itemize}
    The first implication is proved.
  \item For the implication from right to left, we assume that
    $n\not\in\fn{P}$.  We consider $m \neq n$, and some $Q,R$ as in
    the statement of the lemma.  Then
    $n\not\in\fn{\amb{m}{P\,|\,Q}\,|\,R}$, so that for any $T$ such
    that $\amb{m}{P\,|\,Q}\,|\,R~\Rar~T$, $n\not\in\fn{T}$.\Qed
  \end{itemize}
\newcommand{\toplevel}[1]{\copyright^{1} #1}

We can now define the formula $ \copyright n $ to capture the set of
free names of a process, together with the two auxiliary formulas $
\Ftext{flat}$ and $ \toplevel n $ needed in the definition of $
\copyright n $. These formulas are given in
Table~\ref{t:freenamesform}.

\begin{table}[ht]
\myframe{
    \begin{array}{lcl}
    \Ftext{flat} & \eqdef & (\exists x.~\Ftame{\inamb{x}}.\zero~\lor~
    \Ftame{\outamb{x}}.\zero~\lor~\Ftame{\openamb{x}}.\zero~\lor
    \amb{x}{\zero} ~\lor~ \UmsgN x)^\omega
    \\
    \toplevel n & \eqdef &
    \big(\Fmeta{\inamb{n}}.\ltrue~~\lor~~ \Fmeta{\outamb{n}}.\ltrue
    ~~\lor~~\Fmeta{\openamb{n}}.\ltrue~~\lor~~\amb{n}{\ltrue}~~\lor~
    \UmsgN n~\big) 
    ~~\big|~~\ltrue
    \\
    \copyright n & \eqdef & \forall x.~ (\Ftext{flat}\land\non\toplevel n)
    ~\brtr~ \big(~(\Ftext{flat}\land\non\toplevel n)~\brtr~ \sometime~
    \amb{x}{\toplevel n}~\big)@x
  \end{array}}

\caption{Formulas for free names}
\label{t:freenamesform}
\end{table}
Formula $\copyright n$ detects whether name $n$ occurs in a
process, while $\toplevel n$ detects whether $n$ occurs
at top level (i.e. $P$ satisfies this formula iff $\deep_n(P)=0$).

\begin{thm}[Name occurrence]
  $P~\sat~\copyright n$ iff $n\in\fn{P}$.
\end{thm}
\proof
  Consequence of the previous lemma.\Qed

\subsection{Formulas for persistence}\label{subsec:persistence}

We now move to the definition of formulas that characterise
\emph{persistence}, which is given by the replication operator in MA. 
In other words, we investigate the possibility of 
 defining formulas $!\AAA$ that detect replicated term $!P$ such that
 $P$ satisfies $\AAA$. 
However, we cannot hope to define  arbitrary formulas with precisely
this property. First, the form $!P$ is too restrictive: as $P\eqL Q$ implies $!P\eqL !P|Q$ (see \cite{Part2}), a formula $!\AAA$ would not distinguish between a uniquely replicated process $!P$, and a replicated process "with admissible garbage" $!P|Q$ or $!P|!Q$. Second, if we want to express that the process holds
something replicated, one has to reject formulas satisfied by the process $\nil$. 

We hence restrict our attention to the case of formulas $\A$ whose models are single processes only. For these formulas, $!\A$ characterizes replicated processes, in the sense  that
$$
P\sat!\A \quad\Leftrightarrow\quad \exists P_1,\dots,P_n~\mbox{s.t.}~\left\{\begin{array}{l}
1)~~ P\equiv !P_1|(!)P_2|\dots|(!)P_n \\ 2)~~\forall i\in1\dots n,~P_i\sat\A 
\end{array}\right.
$$
where $(!)$ denotes an optional replication.
In the sequel, we show how to define the formula $!\A$ when $\A$
characterizes a guarded process and has some extra conditions. For the
purpose of defining characteristic formulas, this will be
sufficient. However, it remains  an open question how 
to define $!\A$ on a larger language.

The definition of $!\AAA$ has two parts.  The first part says that if
$P \sat !\AAA$ then all parallel components in $P$ that are single and
at top level satisfy $ \AAA $.  This is expressed by the formula
$\AAA^{\omega}$.  The second part of the definition of $!\AAA$
addresses persistence, by saying that there are infinitely many
processes at top level that satisfy $\AAA$ in the sense that we may
not consume all copies by some finite sequence of reduction.
Definitions are given in Table~\ref{t:persistency}: there is one
formula for each possible topmost constructor (recall that we are
considering a single process).

\begin{table}[t]
\myframe{
  \begin{array}{lcl}
    \repl{\inamb{n}}{\AAA} & \eqdef &
    ~~\AAA^{\omega}
    ~~\land~~\forall m.~(\non\copyright m)~\to\\
    &&(\Ftame{\outamb{n}}.\zero)^{\omega}~
    \rtr~\big(~\amb{n}{\zero}  
    ~\rtr~\square\sometime~(~\amb{n}{\amb{m}{\AAA~|~\ltrue}}
    ~)\big)@m
    \\
    \repl{\outamb{n}}{\AAA} & \eqdef &
    ~~\AAA^{\omega}~~\land~\forall
    m.~(\non\copyright m)~\rightarrow\\ 
    &&(\Ftame{\inamb{n}}.\zero)^{\omega}~\rtr~\big(~\amb{n}{\zero}
    ~\rtr~\square~\sometime(~\amb{m}{\AAA~|~\ltrue}|\amb{n}{\zero}
    ~)\big)@m
    \\
    \repl{\openamb{n}}{\AAA} & \eqdef &
    ~~\AAA^{\omega}~\land~(\amb{n}{\zero})^{\omega}~\rtr~\square~
    (~\AAA~|~\ltrue~)
    \\
    \repl{\amb{n}{}}{\AAA} & \eqdef & ~~(\amb{n}{\AAA})^{\omega}~
    \land~(\Ftame{\openamb{n}}.\zero)^{\omega}~\rtr~
    \square~(~\amb{n}{\AAA}~|~\ltrue~)
    \\
    \F_{!\msg n} & \eqdef & {\UmsgN n}^\omega ~\land~
    \TestComm^\omega ~\rtr~ \square~(~\UmsgN n~|~\ltrue~)
    \\
    \repl{input}{\AAA} & \eqdef & \AAA^\omega ~\land~ \mathtt{1mess}^\omega
    ~\rtr~ \square~ (~\AAA~|~\ltrue~)
    \\ 
   \end{array}}
\caption{Formulas for persistent single terms}
\label{t:persistency}
\end{table}

Formula $\F_{!\msg n}$ is actually a characteristic formula, since it
is satisfied only by the process $!\msg n$. For this reason, we
anticipate the notation $\F_P$ of the characteristic formula of $P$
(see Section~\ref{s:chara}).
 For the other formulas, we express the
replication of a process satisfying $\A$; the interpretation of these
formulas hence relies on the actual meaning of $\A$.  

To illustrate the point, consider formula $\repl{\openamb
  n}{\may{\openamb n}.\ltrue}$. This formula only specifies that any
number of capabilities $\openamb n$ should be present at top-level, and
thus holds for process $!\openamb n.\nil$, but also for $\openamb
n.!\openamb n.\nil$.  On the other hand, $\may{\openamb n}.\ltrue$ can
be replaced by the more discriminating formula $\must{\openamb
  n}.\zero$: then we obtain a formula that only accepts process
$!\openamb n.\nil$.

In light of these observations, we define the following measures on
terms:

\begin{definition}[Sequentiality degree, $\ds$]\label{defds}
  The sequentiality degree of an open term is defined as follows:
  \begin{itemize}
  \item $\ds(\nil)=0$, $\ds(P|Q)=\max\big(\ds(P),\ds(Q)\big)$;
  \item $\ds(\amb{\eta}{P})=\ds(!P)=\ds(P)$;
  \item $\ds(\capa.P)=\ds(P)+1$.
  \item $\ds(\msg \eta)=1$ and $\ds((x)P)=\ds(\downarrow_\beta P)+1$ 
    \end{itemize}
\end{definition}

\begin{definition}[Depth degree] 
  The depth degree of a process is given by a function
  $\dd$ from $\MA$ processes to natural numbers, inductively
  defined by: 
  \begin{itemize}
  \item $\dd(\nil)~=~ 0$, $\dd(\capa.P)=\dd((x)P)=\dd(\msg \eta)~=~ 0$;
  \item $ \dd(\amb{\eta}{P})~=~\dd(P)+1 $;
  \item $\dd((!)P_1|\dots|(!)P_r)~=~\max_{1\leq i\leq r}
    \dd(P_i)$.
  \end{itemize}
\end{definition}

\begin{lem}\label{measureequiv}
  For any processes $P$ and $Q$, $P\equiv Q$ implies $\ds(P)=\ds(Q)$
  and $\dd(P)=\dd(Q)$.
\end{lem}


\begin{definition}[Selective and expressive formulas]
 A formula is \emph{sequentially} (resp.\  \emph{depth})
  \emph{selective} if all processes satisfying it have the same
  sequentiality (resp.\ depth) degree. 
  
  For any capability $\capa$ (resp.\ name $n$) and formula $\A$, $\AAA$
  is \emph{$\capa$-expressive} (resp.\ \emph{${n}{}$-expressive,
    input-expressive}) if all terms satisfying it are of the form
  $\capa.P$ (resp.\  $\amb{n}{P}$,$(x)P$).
\end{definition}

\begin{example}
  $\may{\inamb n}.n[\zero]$ is $\inamb n$-expressive but not
  sequentially selective: it admits both $\inamb n. n[\nil]$ and
  $\inamb n.(n[\nil]|\openamb n.n[\nil])$ as models.  On the other
  hand, $\must{\inamb n}.\amb{n}{\zero}$ is both sequentially
  selective and $\inamb n$-expressive.  As we will see below
  (Subsection~\ref{subsec:maif}), the combination of $\may{\capa}$ and
  $\must{\capa}$ modalities allows us to define sequentially selective
  formulas.
\end{example}

These two forms of selectivity are useful for the characterisation of
persistence. Indeed, the sequentiality (resp.\ depth) degree of a
single prefixed (resp.\ ambient) term is strictly decreasing when
consuming the prefix (resp.\ opening the ambient). This property is
needed in order to detect the presence of replication at top-level in a
process, and interpret the formulas introduced above.

\begin{lemma}\label{dsred}
Let $P,Q$ be two terms of \MA{}. If $P\,\rar\, Q$ or $P\,\arr{\mu}\, Q$ 
for some $\mu$, then $\ds(P) \geq \ds(Q)$.
\end{lemma}

\proof
  The property for $\arr{\mu}$ follows from the definition of
  $\ds(P)$.  For $P\,\rar\, Q$, one reasons by induction and case
  analysis (using Lemma~\ref{lem:basic-reduction}).\Qed

\begin{coro}
For all $\capa$, if $P\Rcap Q$, then $\ds(P)\geq \ds(Q)$.
\end{coro}

In the sequel,  $\Pi_{1\leq i \leq t } Q_i$ abbreviates $Q_1 |
    \ldots | Q_t$.

\begin{lem}[Characterisation of replication of single processes]
\label{intcaplem2}\mbox{}
\begin{enumerate}
\item Given a capability $\capa$, and a sequentially selective and
  $\capa$-expressive formula $\AAA$, define
  $$!\AAA~\defiDS~\repl{\capa}{\AAA}.$$ Then $P \sat!  \AAA$ iff there
  are $r\geq 1, s \geq r, P_i$ ($1\leq i\leq s$) such that 
\[P \equiv
  \Pi_{1\leq i \leq r} !{\capa.P_i}\,|\, \Pi_{r+1\leq i \leq s}
  {\capa.P_i}
  \qquad\hbox{and}\qquad
  \capa.P_i\,\sat\,\AAA
  \quad\hbox{for all $1\leq i\leq s$.}
\]
  \item For any name $n$ and depth selective and
    ${n}{}$-expressive formula $\AAA$, define
    $$!\AAA~\defiDS~\repl{\amb{n}{}}{\AAA}.$$  Then $P \sat! \AAA$ iff
    there are $r\geq 1, s \geq r, P_i$ ($1\leq i\leq s$) such that 
\[P\equiv \Pi_{1\leq i \leq r} !{\amb n{P_i}}\,|\, \Pi_{r+1\leq i
      \leq s} {\amb n{P_i}}
  \qquad\hbox{and}\qquad
  \amb n{P_i}\,\sat\,\AAA
  \quad\hbox{for all $1\leq i\leq s$.}
\]
  \item For any formula $\A$ that is sequentially selective and input
    expressive, define $$!\A~\defiDS~\repl{input}{\A}.$$ Then $P\sat
    !\A$ iff there are $r\geq 1, s \geq r, P_i$ ($1\leq i\leq s$) such
    that 
\[P\equiv \Pi_{1\leq i \leq r} !{(x)P_i}\,|\,
  \Pi_{r+1\leq i \leq s} {(x)P_i}
  \qquad\hbox{and}\qquad
  (x) P_i\,\sat\,\AAA
  \quad\hbox{for all $1\leq i\leq s$.}
\]
\end{enumerate}
\end{lem}

\proof
Let us examine some cases:

\noindent\textbf{Case 1, $\capa=\inamb{n}$.} Assume there exist some
terms $P_1, \dots, P_s$ satisfying the condition expressed in $1$.
Then the first part of $!\AAA$ is satisfied, i.e. $P \sat
\AAA^\omega$.

To establish the second part, we have to show that for any
$Q\,\equiv\,\outamb{n}^{\omega}$ (where
$\omega\in\N^*\cup\{\infty\}$), any fresh name $m$, and any term $R$
such that $\amb{m}{P\,|\,Q}\,|\,\amb{n}{\nil}~\Rar~R$, there is a
further reduction $R\,\Rar\,\amb{n}{\amb{m}{R_1\,|\,R_2}}$ for some
$R_1, R_2$ such that $R_1\sat\AAA$, which entails in particular
$R_1\,\equiv\,\inamb{n}.R'_1$.
Since ambient $n$ does not contain any active process, and since there
is no active process at top-level in
$\amb{m}{P\,|\,Q}\,|\,\amb{n}{\nil}$, ambient $n$ remains at top-level
in all evolutions of this term. Moreover, we have that $m$ is fresh
for $P$ and $Q$; therefore, no ambient may get out of $m$, so for any
reduct $R$, there exists $R'$ such that either $(i)$
$R\,\equiv\,\amb{m}{R'}\,|\,\amb{n}{\nil}$, and
$P\,|\,Q~\Stat{\inamb{n}}{\outamb{n}}\, R'$, or $(ii)$
$R\,\equiv\,\amb{n}{\amb{m}{R'}}$, and
$P\,|\,Q\,\redin{n}\Stat{\outamb{n}}{\inamb{n}}\,R'$. In the first
case, because of the shape of $P$, we may perform one more step of
reduction to reach a situation like $(ii)$, and then, since
$P\,|\,Q\,\redin{n}\Stat{\outamb{n}}{\inamb{n}}\, R'$, there exists
$R''$ such that $R'\,\equiv\,!\inamb{n}.P_1\,|\,R''$. The first
implication is thus proved.
  
Conversely, let us assume that $P~\sat~\repl{\inamb{n}}{\AAA}$.  Then
according to the first part of the formula, there exist some $P_i$'s
satisfying
$P\,\equiv\,(!)\inamb{n}.{P_1}\,|\dots|\,(!)\inamb{n}.{P_r}$ and
$\inamb{n}.P_i~\sat~\AAA$. Suppose now by absurd that no component is
replicated. We exploit the sequential selectivity hypothesis to obtain
a contradiction. Indeed, we have the reduction
$\amb{m}{P\,|\,(\outamb{n})^r}\,|\,\amb{n}{\nil}~
\Rar~R=\amb{n}{\amb{m}{P_1\,|\dots|\,P_r}}$ and $R$
is a term whose sequentiality degree is strictly smaller than $\ds(P)$.
Then it is also the case for any of its reducts, and therefore the
same reasoning holds for any $R_1, R_2$ such that
$R~\Rar~\amb{n}{\amb{m}{\inamb{n}.R_1\,|\,R_2}}$, $\inamb{n}.R_1$ has
a sequentiality degree too small to satisfy $\AAA$ because of
sequential selectivity.  Thus, $P$ cannot satisfy
$\repl{\inamb{n}}{\AAA}$, and we obtain a contradiction. Hence, at
least one of the $P_i$'s 
is replicated, and the reverse implication is
proved.

\noindent The proofs for \textbf{Case 1}, other capabilities, and
\textbf{Case 3} follow from similar arguments.

\smallskip
\noindent\textbf{Case 2.}  Assume that
$P\equiv!\amb{n}{P_1}\,|\dots|\,(!)\amb{n}{P_r}$, with the $P_i$'s such
that $P_i~\sat~\AAA$. Then $P$ satisfies $\repl{\amb{n}{}}{\AAA}$ iff
for any $Q\equiv\openamb{n}^{\omega}$, and any $R$ such that
$P\,|\,Q\Rar R$, there are $R_i$'s such that
$R\equiv\amb{n}{R_1}\,|\,R_2$ with $\amb{n}{R_1}\sat\AAA$.  Since for
any $R$, $R\equiv!\amb{n}{P_1}\,|\,R'$, the first implication is
established.
  
Conversely, suppose $P$ satisfies $\repl{\amb{n}{}}{\AAA}$.  Then
$P\equiv(!)\amb{n}{P_1}\,|\dots|\,(!)\amb{n}{P_r}$. Moreover, if no
$P_i$ is replicated,
$P\,|\,!\openamb{n}~\Rar~P_1\,|\dots|\,P_r\,|\,!\openamb{n}$, and if
in some $P_i$ there are $P_{i,j}$ ($j=1,2$) such that
$P_i\equiv\amb{n}{P_{i,1}}\,|\,P_{i,2}$, then the depth degree of
$P_{i,1}$ is too small for $\amb{n}{P_{i,1}}$ to satisfy $\AAA$, which
gives us the second implication.\Qed

The formulas for persistence, together with the constructions of
Section~\ref{s:capacom}, will be used to derive characteristic
formulas with respect to $\eqL$ for a sub-calculus of MA in
Section~\ref{s:chara}.


\section{Characteristic formulas}\label{s:chara}

In this section we establish the existence of characteristic formulas
for a large class of processes. Given a process $P$, a characteristic
formula for $P$ is a formula $\mathcal{F}_{P}$ such that:
$$
\forall Q.~~~Q~\sat~\mathcal{F}_P~~~\mbox{iff}~~Q\,\eqL\, P\,,
$$
\noindent where $\eqL$ is logical equivalence (i.e., $P\,\eqL\,Q$ iff
$P$ and $Q$ satisfy the same formulas).

The definability of characteristic formulas is an interesting property, though for now only a purely theoretical result. The effectiveness and efficiency of the construction of characteristic formula are beyond the scope of this
paper, though we strongly believe that our definition gives an algorithm for constructing
formulas on the semi-decidable fragment $\MAIF$. Having such constructive characteristic formulas,
would have some practical impact, since
we could relate the logical equivalence and model-checking problem to the validity problem. Interestingly, we may also recall that validity reduces to model-checking the other way round
when the spatial logic considered has the guarantee ($\rtr$) connective.

To be able to carry out our programme, we have first to understand
what $\eqL$ represents.  For this, we use a co-inductive
characterisation of $\eqL$, as a form of labelled bisimilarity. Then,
making an intensive use of the formulas for the connectives of the
calculus previously defined, we derive the characteristic formulas.

\subsection{Intensional bisimilarity}\label{subsec:intbis}
\mbox{}\\
\underline{ Note for this subsection only.}  The results presented in this
subsection have appeared previously
in~\cite{Sangiorgi::ExtInt::01,Sedal} and therefore are not a
contribution of the present paper. Their complete proofs, which are
rather long and complex, can be found in a companion
paper~\cite{Part2}.
We will use the notion of intensional bisimilarity
and all the properties that are recalled in this subsection only in the proof
about characteristic formulas for AL (Theorem~\ref{thm:charac}), which is one of our main
expressiveness results.

\medskip

We use the labelled transitions (Definition~\ref{d:statt}) to define a
notion of intensional bisimilarity in order to capture $\eqL$.

\begin{definition}
\label{d:bisMOD}
\emph{Intensional bisimilarity} is the largest symmetric relation
$\bisMOD$ on closed processes such that $P \bisMOD Q$ implies:
\begin{enumerate}
\item
\label{cMOD:par}
If $P \equiv P_1| P_2$ 
then there are $Q_1,Q_2$ such that 
$Q \equiv  Q_1|Q_2 $ and
$P_i \bisMOD Q_i$, for $i=1,2$.

\item
\label{c:nil}
 If $P \equiv \nil $ then  $Q \equiv \nil$.

\item
\label{c:tau}
 If $P \longrightarrow P'$ then there is $Q'$ such that 
$Q\Longrightarrow Q'$ and $P' \bisMOD Q'$.
\item
\label{cMOD:in}
 If 
$P \arr{\ina n} P'$
 then 
there is $Q'$ such that
 $Q \Ar{\ina n } \StatOI n Q' $
and $P' \bisMOD Q'$.

\item
\label{cMOD:out}
 If $P \arr{\out n } P'$ then 
there is $Q'$ such that 
$Q \Ar{\out n } \StatIO n Q' $
and $P' \bisMOD Q'$.

 \item
\label{cMOD:open}
 If $P\arr{ \open n } P'$ then 
there is $Q'$ such that $Q \Ar{\open n } Q' $
and $P' \bisMOD Q'$.

\item
\label{cMOD:msg}
 If $P \arr{\msg n} P'$ then 
there is $Q'$ such that 
$Q \Ar{\msg n}  Q' $ and $P' \bisMOD Q'$.

\item
\label{c:absMOD} If $P   \arr{? n}  {P'}$
then
there is $Q'$ such that $Q | \msg n \Longrightarrow Q'
$ and  $P' \bisMOD Q'$.

\item 
\label{cMOD:amb}
If $P \equiv \amb n {P'}$ then there is $Q'$ such that $Q 
\equiv \amb n {Q'}$ and $P' \bisMOD Q'$.
\end{enumerate} 
\end{definition}

The definition of $\bisMOD$ has (at least) two intensional clauses,
namely \reff{cMOD:par} and \reff{c:nil}, which allow us to observe
parallel compositions and the terminated process. These clauses
correspond to the intensional connectives `$|$' and `$\zero$' of the
logic.  The clause \reff{c:absMOD} for abstraction is similar to the
input clause of bisimilarity in asynchronous message-passing calculi
\cite{AmCaSa98}. This is the case because communication in MA is
asynchronous. Another consequence of this is that the logic
is insensitive to the following rewrite rule (modulo
associativity-commutativity of $|$):
$$
(x)\big(\msg x|(x)P\big)~~\rar_\eta~~ (x)P\,.
$$
This rule induces a notion of normal form of processes, that we
shall call the \emph{eta-normalised form}. 

\begin{definition}[Eta-equivalence]\label{d:etaeq}
We will note $P\equivE Q$ if the normal forms of $P$ and $Q$ for 
$\rar_{\eta}$ are related by $\equiv$.
\end{definition}

\begin{lem}[\cite{Sedal,Part2}]\label{lem:etaeqL}
  For any closed process $P,Q$ in \MA{}, $P\equivE Q$ implies
  $P\bisMOD Q$.
\end{lem}

By Theorem~\ref{p:correction} below, this result says that the logic
is insensitive to $\rar_{\eta}$. We shall thus reason using normalised
processes with respect to $\rar_\eta$ in the proof of Theorem~\ref{thm:charac}.

The most peculiar aspect of the definition of $\bisMOD$ is the use of
the stuttering relations.  Although they can be avoided on finite
processes, they cannot in the full calculus.
By contrast, stuttering does not show up in Safe
Ambients~\cite{LeSa00full}, where movements are achieved by means of
synchronisations involving a capability and a \emph{co-capability}.

We now state some results about $\bisMOD$ that are proved
in~\cite{Sedal,Part2}.

\begin{thm}
  \label{p:correction}
  For any $P$, $Q$, $P\bisMOD Q$ implies $P\,\eqL\, Q$.
\end{thm}

The latter result establishes \emph{correctness} of $\bisMOD$ with respect to
$\eqL$. 
Given a process $P$, we try and characterise the equivalence class of
$P$ with respect to $\bisMOD$ with a formula $\mathcal{F}_P$.
The definability of such a formula will actually entail that
$\eqL~\subseteq~\bisMOD$ (\emph{completeness}), and hence that
$\mathcal{F}_P$ actually characterises the $\eqL$-equivalence class of
$P$.

We now mention a useful induction principle that allows us to reason
\emph{`almost inductively'} on the structure of a process when
checking relation $\bisMOD$. This principle is given by the following
inductive order:

\begin{definition}\label{def:order}
  We write $P>Q$ if either $\ds{(P)}>\ds{(Q)}$ or $Q$ is a sub-term of
  $P$. 
\end{definition}

This order allows us, using the following result, to derive an
inductive characterisation of $\bisMOD$~\cite{Sedal,Part2}.

\begin{prop}
  \label{p:inversion}
  Let $P, P_1, P_2, Q$ be processes of \MA.  Then
\begin{enumerate}
\item $\nil~\bisMOD~Q$ iff $Q\equiv \nil$.
\item $\amb{n}{P}~\bisMOD~Q$ iff there exists $Q'$ such that
  $Q~\equiv~\amb{n}{Q'}$ and $P~\bisMOD~Q'$.
\item $P_1|P_2~\bisMOD~Q$ iff there exist $Q_1,Q_2$ such that
  $Q~\equiv~Q_1|Q_2$ and $P_i~\bisMOD~Q_i$ for $i=1,2$.
\item $!P~\bisMOD~Q$ iff there exist $r\geq 1, s \geq r, Q_i$ 
  ($1\leq i\leq s$) such that $P~\bisMOD~Q_i$ for $i=1\dots s$, and 
  $Q \equiv \Pi_{1\leq i \leq r} {!Q_i}\,|\, \Pi_{r+1\leq i \leq s}
  {Q_i}$.
\item $\capa.P~\bisMOD~Q$ iff there exists $Q'$ such that
  $Q~\equiv~\capa.Q'$ with $P~\Rcap\bisMOD~Q'$ and
  $Q'~\Rcap\bisMOD~P$.
\item $\msg n~\bisMOD~Q$ iff $Q\equiv \msg n$.
\item 
$(x)P~\bisMOD~Q$ iff there exists $P',Q',Q''$ and 
$n\not\in\fn{P}\cup\fn{Q}$ such that $Q\equiv(x)Q'$,
  $Q|\msg n\Rar Q''$, $\downarrow_\beta P\sub n x \bisMOD Q''$, $(x)P|\msg n\Rar P'$
  and $P'\bisMOD Q'\sub n x$.

\end{enumerate}
\end{prop}

\subsection{The sub-calculus \MAIF}\label{subsec:maif}

As we mentioned above, characteristic formulas and completeness for an
algebraic characterisation of logical equivalence are two related
problems.  In fact, the existence of characteristic formulas is a
stronger result than completeness of $\bisMOD$ with respect to $\eqL$: while we
establish completeness in~\cite{Part2} on the whole calculus, we are
only able to derive characteristic formulas on a sub-calculus of MA. 
To introduce the necessity of restricting the class of processes we
consider, and to illustrate the basic ideas behind the construction of
characteristic formulas, we examine some examples.

\begin{example}
  
  We introduce the following processes: $P_1=!  \openamb n .
  \amb n \nil$, $P_2=\openamb n | \amb{n}{\nil}$,
  $P_3=!\openamb{n}.P_2$, and $P_4 = \openamb n.P_2$.
  
  A characteristic formula for $P_1$ is easy to define since the
  continuation term $\amb{n}{\nil}$ has no reducts. Hence the formula
  $\cn {\openamb n} \amb n \zero$, using a formula for
  \emph{necessity}, satisfies the conditions of
  Lemma~\ref{intcaplem2}, and a characteristic formula for $P_1$ is
  $$\F_1~\defiDS~! \cn {\openamb n} \amb n \zero\,.$$
  In order to define a characteristic formula for $P_3$, we first look
  for a characteristic formula for $P_2$. We can set
  $$
  \F_2~\eqdef~ \Ftame{\openamb{n}}.\zero \,|\,\amb{n}{\zero}\,.
  $$
  $\F_2$ is indeed a characteristic formula for $P_2$. However,
  $\Ftame{\openamb{n}}.\F_2$ is not a characteristic formula for
  $P_4$, nor for $P_3$. The reason is that the continuation process
  ($P_2$) is not static, as it may reduce to $\nil$. Hence
  $\Ftame{\openamb{n}}.\F_2$ does not satisfy the conditions of
  Lemma~\ref{intcaplem2}, so that we need to add the possibility to
  reduce to $\nil$, yielding the formula
  $\Ftame{\openamb{n}}.(\F_2~\lor~\zero)$. But then we also accept the
  term $\openamb{n}.\nil$, which shows why we are led to add a
  \emph{possibility} condition to the formula, and we finally define
  the following characteristic formula for $P_3$:
  $$
  \F_3~\eqdef~\repl{\openamb{n}}{\Fmeta{\openamb{n}}.\F_2~\land
    ~\Ftame{\openamb{n}}.(\F_2~\lor~\zero)}\,.
  $$
\end{example}

We see on this example that characterising the continuation of a
process starting with a capability or an input requires to enumerate
also all the possible reducts after consuming the topmost constructor.
Therefore, the definition of characteristic formulas relies on the
actual feasibility of such an enumeration, which leads us to the
definition of a subclass of MA processes.

In the definition below, we use the following notation: given a set
$\cal{S}$ of processes, $\cal{S}_{/\bisMOD}$ will stand for the
quotient of $\cal{S}$ with respect to $\bisMOD$ (which is,
technically, a set of $\bisMOD$-equivalence classes of processes).

\begin{definition}[Sub-calculus \MAIF]
  A process $P$ is \emph{image-finite} iff any sub-term of $P$ of the
  form $\capa.P'$ (resp. $(x)P'$) is such that the set $ \{
  P''~:~P'~\Rcap~P''\}_{/\bisMOD}$ (resp. $\{ P''~:~P'\sub n
  x~\Rar~P''\}_{/\bisMOD}$, for some $n\not\in\fn{P}$) is finite.

  \MAIF{} is the set of image-finite MA processes.
\end{definition}

\MAIF{} is only a semi-decidable fragment of MA. A stronger
restriction is considered in \cite{Part2}, whose definition involves
decidable syntactic conditions. We however stick to this larger
fragment for the sake of generality.

%

For example, process $\inamb n.!(n[\nil]|\openamb n.\nil)$ is in
\MAIF{}, but $\inamb n.!(n[\nil]|\openamb n.a[\nil])$ is not.

To construct a characteristic formula $\F_P$ for a closed \MAIF{}
process $P$, we can suppose (up to $\equivE$) that replication only
appears above single terms and that $P$ is eta normalised.  We then
define the characteristic formula $\F_P$ of $P$ by induction using the
order of Definition~\ref{def:order} (this defines a valid induction by
Lemma~\ref{dsred}). The defining formulas are given in
Table~\ref{t:charactform}. Two technical remarks should be made
regarding the definition of $\F_{(x)P}$. First, in the disjunction
over the quotiented set $\{P':~P\sub {n_x} x \Rar P'\}_{/\bisMOD}$, it
is intended that we pick a representative in each equivalence
class. Second, to avoid reasoning about processes containing free
variables (characteristic formulas are defined only for closed
processes), we introduce the auxiliary name $n_x$, that is used as a
placeholder for $x$, to be replaced by $x$ again once the
characteristic formula of process $P'$ has been computed (see the
defining clause of $\F_{(x)P}$). So $\F_{P'}\sub x {n_x}$ is a slight
abuse of notation that denotes the operation consisting of $(i)$
alpha-converting $\F_{P'}$ so that no bound variable is named $x$, and
$(ii)$ textually replacing $n_x$ with $x$ in the resulting formula.

\begin{table}[ht]
%
%
$$\boxed{
\begin{array}{lclclcl}
  \!\F_{\mbox{\begin{footnotesize}\nil\end{footnotesize}}} & \defiDS &
  \zero 
  &&
  \F_{P|Q} & \defiDS &\F_{P}~|~\F_{Q}~~~~
  \\
  \!\F_{\amb{n}{P}} & \defiDS & \amb{n}{\F_P}
  &&
  \F_{\mbox{\begin{footnotesize}\capa\end{footnotesize}}.P} & \defiDS &
  \Fmeta{\capa}.\F_P~\land~\Ftame{\capa}.\bigvee_{\{P',~P\Rcap
    P'\}_{/\bisMOD}} \F_{P'}
  \\
  \!\F_{!\amb{n}{P}} &
  \defiDS & 
  \repl{\amb{n}{}}{\F_P}
  &&
  \F_{!\mbox{\begin{footnotesize}\capa.\end{footnotesize}}P} & \defiDS
  & \repl{\capa}{\F_{\mbox{\begin{footnotesize}\capa\end{footnotesize}}.P}} 
  \\[.6em]
  \!\F_{\begin{footnotesize}(x)P\end{footnotesize}} & \defiDS & 
  \multicolumn{5}{l}{
    \exists x.~\non \copyright x~\land~\may {?x}.
    \mathcal{F}_{P}~\land~
    \qquad\qquad\qquad\qquad (n_x\notin\fn{P})
    }
  \\
  & &
  \multicolumn{5}{l}{
  \must{? x}\Big(\big(\F_{\msg
  {x}}|(\mathtt{1input}\land\non\copyright x)\big)~\lor~\bigvee_{\{P':~P\{ {n_x}/ x\} \Rar 
    P'\}_{/\bisMOD}} \F_{P'}\sub x {n_x}\Big)
  }
  \\[.6em]
  \!\F_{\begin{footnotesize}\msg n\end{footnotesize}} &\defiDS 
  &\mbox{cf. Lemma~\ref{l:UcommNN}}
  &&
    \F_{\begin{footnotesize}!\msg n\end{footnotesize}} & \defiDS 
  &
  \mbox{cf. Table~\ref{t:persistency}} 
  \qquad\quad
  \F_{\begin{footnotesize}!(x)P\end{footnotesize}}~\,\defiDS\,~
  \repl{input}{\F_{(x)P}}\!
  \\
\end{array}}
$$
\caption{Characteristic formulas in \MAIF}
\label{t:charactform}
\end{table}


\begin{theo}[Characteristic formulas for $\Pb$]\label{thm:charac}
  For any closed term $P$, define $\F_P$ according to
  Table~\ref{t:charactform}. Then\\
  \mbox{}\hfill$Q~\sat~\F_{P}\qquad\mbox{iff}\qquad
  P~\bisMOD~Q$.\hfill\mbox{}
\end{theo}

\proof 
  The proof is by induction, using the order of
  Definition~\ref{def:order}.
\begin{itemize}
\item $\F_{\nil}$ characterises $\nil$: this holds by
  Proposition~\ref{p:inversion}.
\item $\F_{\msg n}$ characterises $\msg n$ and $\F_{!\msg n}$ characterises 
$!\msg n$: by Lemma~\ref{l:UcommNN}, Lemma~\ref{intcaplem2} and 
Proposition~\ref{p:inversion}.
\item if $\F_{P}$ characterises $P$, then $\F_{n[P]}$ characterises $n[P]$:
 by Proposition~\ref{p:inversion}.
\item if $\F_{P_1}$ characterises $P_1$ and $\F_{P_2}$ characterises
  $P_2$, then $\F_{P_1|P_2}$ characterises $P_1|P_2$: by
  Proposition~\ref{p:inversion}.
\item Suppose now that for every $P'$ such that $\ds(P')\leq \ds(P)$,
  $\F_{P'}$ is a characteristic formula for $P'$. We then have:
\begin{itemize}
\item $\F_{\capa.P}$ characterises $\capa.P$.
  
  By Lemma~\ref{dsred}, $\ds(P')\leq \ds(P)$ for any $P'$ such
  that $P\Rcap P'$, so $\F_{P'}$ is a characteristic formula for such
  processes.  We examine each of the two implications. In one
  direction, $\capa.P\sat \may\capa.\F_P$, and by Lemma~\ref{capacformul},
  $\capa.P\sat\must\capa.\bigvee_{\{P',~P\Rcap P'\}_{/\bisMOD}} \F_{P'}$, so
  $\capa.P\sat\F_{\capa.P}$.  Conversely, if $Q\sat\F_P$, then from
  $Q\sat\may\capa.\F_P$ we deduce the existence of $Q',Q''$ such that
  $Q\equiv \capa.Q'$, $Q'\Rcap Q''$, and $Q''\sat \F_P$. Moreover,
  from $Q\sat\must\capa.\bigvee_{\{P',~P\Rcap P'\}_{/\bisMOD}}
  \F_{P'}$, we deduce that there is $P'$ such that $P\Rcap P'$ and
  $Q'\sat F_{P'}$, so $Q'\bisMOD P'$, and by Proposition~\ref{p:inversion},
  $Q\bisMOD \capa.P$.
\item $\F_{(x)P}$ characterises $(x)P$.
  
  We first prove that $(x)P\sat\F_{(x)P}$.  We pick $n_0$ fresh for
  $P$. We can apply the induction hypothesis for $P\sub {n_0} x$ and
  for all of its reducts $P'$.  Then the implication from right to
  left follows from Lemma~\ref{l:inputformulas}.
  
  For the other direction, let $Q$ be such that $Q\sat\F_{(x)P}$. We
  assume first that $Q$ is eta normalised.  Let $n_0$ be a name that
  can be used to satisfy formula $\F_{(x)P}$.  Then
  $n_0\not\in\fn{Q}$, and there are $Q',Q''$ such that $Q\equiv
  (x)Q'$, $\msg {n_0}|(x)Q'\Rar Q''$, and $Q''\sat \F_{P\sub {n_0}
    x}$, that is, by hypothesis, $Q''\bisMOD P\sub {n_0} x$.
  Moreover, since $Q$ is eta normalised, $Q'\sub {n_0} x$ is not of
  the form $\msg {n_0} |(x)R$ with $n_0\not\in\fn{(x)R}$, and hence
  this process does not satisfy the formula $(\F_{\msg
    {n_0}}|(\mathtt{1input}\land\non\copyright n_0))$. Therefore,
  there exists $P'$ such that $P\sub {n_0} x\Rar P'$ and $Q'\sub {n_0}
  x\sat\F_{P'}$, that is, by induction, $Q'\sub {n_0} x \bisMOD P'$.
  Using Proposition~\ref{p:inversion}, we deduce $Q\bisMOD (x)P$.
  
  We consider now the case when $Q$ is not eta normalised. Let $Q_0$
  be the eta normal form of $Q$. Then by Lemma~\ref{lem:etaeqL} and
  Theorem~\ref{p:correction}, $Q\eqL Q_0$. Since by hypothesis
  $Q\sat\F_{(x)P}$, $Q_0\sat \F_{(x)P}$ and by the previous arguments,
  $(x)P\bisMOD Q_0$.  Finally, by Lemma~\ref{lem:etaeqL}, $(x)P\bisMOD
  Q$.
  
\item $\F_{!\capa.P}$ characterises $!\capa.P$ and $\F_{!(x)P}$
  characterises $!(x)P$: these results follow from the replication
  case in Proposition~\ref{p:inversion} and from Lemma~\ref{intcaplem2}. In
  particular, the requirements in terms of sequential (or depth)
  selectiveness, and \capa{} (or $n$, input) expressiveness are
  satisfied because the formulas we are using in our constructions are
  characteristic formulas, which, by induction, satisfy such
  requirements.\Qed
\end{itemize}
\end{itemize} 

\begin{cor}
  On the sub-calculus \MAIF, we have $\bisMOD\,=\,\eqL$.
  
  For any closed processes $P$ and $Q$ of \MAIF, we have
\[\hbox to\textwidth{%
  \phantom{\qEd}\hfill$Q~\sat~\F_{P}\qquad\mbox{iff}\qquad P~\eqL~Q\,.$
  \hfill\qEd}
\]
\end{cor}




\section{Extensions of the calculus}\label{s:ext}
In this section, we study extensions of MA with different forms of
communication: we first examine the possibility to emit capabilities
(in addition to names) in messages, and then consider synchronous
communication.  We only show how to capture the modifications brought
to the language, without porting all the constructions seen in the
previous sections. We however believe that our approach would go
through without any major modification.  


We start by pointing out that Lemmas~\ref{l:Uopen},\ref{l:Uint},\ref{l:Uout} about the
interpretation of formulas $\may{\capa}.\A$ hold in the extensions we
consider, since their proofs are insensitive to the presence of
communication in the calculus.

\subsection{Capabilities in messages}

In the original MA calculus~\cite{CaGo98}, messages can also carry
\emph{paths of capabilities}.  To accommodate this in the grammar of
Table~\ref{ta:syn}, all occurrences of $\eta$ are replaced by $M$, and
the path productions
\[M :: = \capa \midd M_1 . M_2 \midd \epsilon\, , \]
are added to those for expressions, 
where $\epsilon $ stands for the empty path.
Thus a capability can be a path, such as $
\open n . \ina m .\open h$. 
Also, the rules
\begin{mathpar}
\epsilon.P ~\equiv~ P
\and
(M_1.M_2). P ~\equiv~ M_1.M_2. P 
\end{mathpar}
are added to those of $\equiv$.  Since messages can now carry names or
capabilities, a type system is introduced~\cite{CaGo99popl} to avoid
run-time errors. We shall assume that all processes are well-typed
(according to the basic Ambient types), which means in particular that
in the interpretation of a formula of the form $\A\rtr\BBB$, processes
that are added in parallel are of the right type. Moreover, we will
say that the argument of an abstraction $(x)P$ is \emph{of capability
  type} whenever the typing ensures that capabilities, and not names,
can be sent to instantiate $x$.

Our main focus will be on the characterisation of these new forms of
messages. For this, we need a formula $\TestCap$, the analogous of the
formula $\TestComm$ of Section~\ref{subsec:comm}, satisfied by all
abstractions that are eta-congruent to $\abs x {\amb m {x.\nil}}$,
where $m$ is some fixed name.

We also need a formula $\pAth M$, for any closed capability $M$, that
identifies those processes that are structurally congruent to $M .
\nil$.  We first discuss an example, namely the formula $\pAth {\ina n
  . \open m }$.  For this, $ \UinN n .\UopenN m . \zero$ is not
enough: this formula is satisfied by $\ina n .\open m . \nil$ but
also, for instance, by processes such as $\ina n . (\msg M | \abs x
{\open m})$, which has some additional I/O, or $\ina n . \out n . \ina
n . \open m .\nil$, which stutters.  A formula $\FFF$ for $\pAth{\ina
  n .\open m }$ could thus be (the actual definition of $\pAth{\ina n
  .\open m }$ will be different; the formula below is easier to read
and semantically equivalent):

\[\begin{array}{rcl}
\FFF& \eqdef &\UinN n  .\UopenN m. \zero  
\\[\tkpPOPL]
&& \andd \; 
\myneg \UinN n . \myneg \Ucomp 
 \\[\tkpPOPL]
&& \andd \; 
 \myneg \may{\ina n} . \may{\out n} .\true
 \\[\tkpPOPL]
&& \andd\; 
\myneg \UinN n .\UopenN m . \myneg \zero 
\end{array}
\]

In the definition of $\FFF$, the second, third and fourth conjuncts
take care of the problems with I/O and stuttering mentioned above.

Here is the complete definition of $\pAth M$ for any path $M$:
\myframe{
\begin{array}{rcl}
\pAth{\open n . M} & \defi & \UopenN n . \pAth M   \\[\tkpPOPL]
&&\andd \; \neg \UopenN n . (\neg \Ucomp \orr \Uamb)  \\[\tkpPOPL]
\pAth{\out n . M} & \defi & \UoutN n. \pAth M  \\[\tkpPOPL]
 &&  \andd \; \neg \UoutN n. (\neg \Ucomp \orr \Uamb)   \\[\tkpPOPL]
 && \andd \; \neg \UoutN n . \UinN n .\UoutN n .\pAth M \\[\tkpPOPL]
\pAth{\ina n . M} & \defi & \UinN n. \pAth M   \\[\tkpPOPL]
 && \andd \; \neg \UinN n . (\neg\Ucomp \orr \Uamb)   \\[\tkpPOPL]
 &&\andd\; \neg \UinN n .  \UoutN n . \UinN n . \pAth M \\[\tkpPOPL]
\pAth{\epsilon . M} & \defi &  \pAth M  \\[\tkpPOPL]
\pAth{\nil} & \defi &  \zero
\end{array} 
}

In the definition of $ \pAth M$, sub-formula $\neg \Ucomp \orr \Uamb$
is used to control process reductions, see Lemma~\ref{l:Ucomp_orr_Uamb}. 

We now define $\TestCap$:
\myframeC{
\begin{array}{rcl}
\TestCap & \defi & 
\Ucomm  \\[\tkpPOPL]
&& 
\andd \; \Ucomm \limp \Box (
\ifpopl
\else
\Dcomm \orr \amb m {\Ucomp})
\fi
 \\[\tkpPOPL]
&&
\andd \; \Ucomm \,\limpE\, \diamond \amb m{\pAth{\ina n} } \\[\tkpPOPL]&&
\andd \; \Ucomm \,\limpE\, \diamond \amb m {{\zero} }  \\[\tkpPOPL]
\multicolumn{3}{l}{\mbox{where $(n,m)$ is any pair of different names. 
} } 
\end{array}
 }
The correctness of this definition is proved along the lines of that
of $\TestComm$.  
The formula $\UmsgN M$, where $M$ is  any closed capability, 
is then 
\myframeC{ \UmsgN M \defi \Ucomm \: \andd \Big(
 \TestCap \limp \diamond \amb m {\pAth M} \Big)
}
\iflong

We now give the key steps that allow us to derive the interpretation
of the formulas presented above.

\begin{lem}
\label{l:Ucomp_orr_Uamb}
Suppose $P \longrightarrow P'$. Then 
$\satDS P {\neg \Ucomp \orr \Uamb}$.\Qed
\end{lem} 
 
\begin{lem}
\label{l:pathCHARACT}
Suppose $M,P$ are closed. Then  $\satDS P {\pAth M}$ iff $P \equiv M .\nil$.
\end{lem}  

\proof
By induction on the size of $M$. If the size is $0$ then $M = \zero$
and  the result follows  easily.  For the inductive case, we proceed
by a case analysis. 
\begin{itemize}
\item $M = \ina n. N$. We have $\satDS P{ \UinN n. \pAth M}$, therefore
by Lemma~\ref{l:Uint}   $P \equiv \ina n . P'$ for some $P'$ such 
that $P' \StatIO n P'' $ and $ \satDS{P''} {\pAth M}$.

However $P'$ cannot stutter, otherwise
  $\satDS P{\UinN n .  \UoutN n
  . \UinN n . \pAth M}$. Also, it cannot be 
$P' \longrightarrow P''' \Longrightarrow P''$ otherwise 
 by Lemma~\ref{l:Ucomp_orr_Uamb}
$\satDS{ P'}{\neg\Ucomp \orr \Uamb}$, hence 
$\satDS P {\UinN n . (\neg \Ucomp \orr \Uamb)}$.
\item $M = \ina n. N$: similar.
\item $M = \open n. N$: similar (without any stuttering phenomenon).

\item $M = \epsilon. N$. In this case, we also have $\satDS P{\pAth N}$, 
  hence by induction $P \equiv N$, hence $P \equiv M$.\Qed
\end{itemize} 

We now adapt the notion of \ambAbs, introduced in
Section~\ref{subsec:comm}, in order to define a class of processes
that will be used to give the interpretation of formula \TestCap.

\begin{definition}[\ambAbs\ and \ambSemiAbs]
\label{d:ambAbs}
The \emph{\ambAbs s} are the subset of processes defined by the
following grammar:
\[  P ::= \abs x {\amb m {x. \nil}} \midd \abs x {(\msg N | P)}\,.\] 
The \emph{\ambSemiAbs s} are the subset of processes defined by the
following grammar:
\[  P ::= \abs x {\amb m {Q}} \midd \abs x {(\msg N | P)}\,\]
where $Q$ is single.
\end{definition}

\begin{lem}
\label{l:mauamCAP}
Given an abstraction $\abs x R$ whose argument is of capability type
and $R$ contains no abstractions,  suppose there are 
messages $M,N$ and substitutions $\Sublz,\Sublpz$ such that
\[
\satDS {\msg M  | \abs x { (R\Sublz )}} {\Box (\Dcomm \orr \amb m \Ucomp)}
\]
and 
\[
\satDS {\msg N  | \abs x { (R\Sublpz )}} {\diamond  \amb m \Ucomp}. 
\]
 Then $\abs x R$ is an \ambSemiAbs\ (i.e., $R \equiv \amb m P$ where
 $P$ is single).\Qed
\end{lem}


\begin{lem}
\label{l:auamCAP}
Given an abstraction $\abs x R$  whose argument is of capability type,

 suppose there are 
messages $M,N$ and substitutions $\Sublz,\Sublpz$ such that
\[
\satDS {\msg M  | \abs x { (R\Sublz )}} {\Box (\Dcomm \orr \amb m \Ucomp)}
\]
and 
\[
\satDS {\msg N  | \abs x { (R\Sublpz )}} {\diamond  \amb m \Ucomp}. 
\]
 Then $\abs x R$ is an \ambSemiAbs.
\end{lem}

\proof
By  induction on the number of nested abstractions in $R$.
 If this number is $0$ then  use 
 Lemma~\ref{l:mauamCAP}.

Suppose the number is greater than $0$.  From 
\[\msg M  | \abs x { (R\Sublz )} \longrightarrow R\Sublz
\sub M x  \]
we derive 
\[ \satDS{R\Sublz \sub M x }{
\Dcomm \orr  \amb m \Ucomp
} \] 
Since $R$ should contain an abstraction, the formula $ \amb m \Ucomp $ is
not satisfied, hence 
\[ \satDS{R\Sublz \sub M x }{
\Dcomm 
} \]  
Using this, the fact that $R$ should contain an abstraction, and
the other judgement in the hypothesis of the lemma  we infer that 
\[ R \equiv \msg{M'} | \abs y Q\]
for some $M',y,Q$. 
This information on $R $ and the judgements in the hypothesis of the lemma imply:
\[
\satDS {\msg{ M ' \Sublz \sub M x} | \abs x { (Q\Sublz \sub M x )}}
 {\Box (\Dcomm \orr   \amb m \Ucomp)}
\]
and 
\[
\satDS{\msg{ M ' \Sublz \sub N x} | \abs x { (Q\Sublz \sub N x )}}
 {\diamond  \amb m \Ucomp}. 
\]
We can now conclude, using the inductive hypothesis on $Q$.\Qed

\begin{lem}
\label{l:auamCAPbis}
Suppose $\abs x R$ is an \ambSemiAbs,  
 whose argument is of capability type, and suppose there are 
messages $M,N$ and substitutions $\Sublz,\Sublpz$ such that
\[
\satDS {\msg M  | \abs x { (R\Sublz )}} {\diamond \amb m {\pAth \ina n}}
\]
and 
\[
\satDS {\msg N  | \abs x { (R\Sublpz )}} {\diamond  \amb m{{\zero}}}. 
\]
 Then $\abs x R$ is an \ambAbs.
\end{lem}

\proof
  By induction on the number of abstractions in $R$. The case when
  this number is $0$ is easy: if $R \not\equiv x.\nil$ then $R$ does
  not satisfy the given formulas.
  
  If the number of abstractions is greater than $0$ then $Q \equiv
  (\msg O | P)$, for some message $O$ and process $P$ and then we
  derive:
\[ \msg M  | \abs x { ( O \Sublz | \abs y {P  \Sublz } )} \longrightarrow 
\satDS{  O \Sublz \sub M x  | \abs y {P  \Sublz \sub M x  } }{\diamond \amb m {\pAth \ina n}} \]
and similarly
\[\satDS{  O \Sublpz \sub N x  | \abs y {P  \Sublpz \sub N x  }
  }{\diamond  \amb m{{\zero}}}  \]
and then we conclude using induction.\Qed

 We say that an  ambient abstraction $P$ is \emph{simple}
if $P =_\beta \abs x {\amb  m {x .\zero}}$ 
where $=_\beta$ is the least congruence that is closed under the rule 
\[ \msg M |  \abs x P  = P \sub M x\,. \]

\begin{lem}
\label{l:auamSIMPLECAP}
Suppose  $\abs x Q$ is an \ambAbs, and that we have
\[
\satDS { \abs xQ} {\Ucomm \,\limpE\, \diamond \amb m {\pAth \ina n}}
\quad\mbox{and}\quad
\satDS { \abs xQ } {\Ucomm \,\limpE\,  \diamond  \amb m{{\zero}}}\,.
\]
 Then $\abs x Q$ is simple.
\end{lem} 

\proof 
Any \ambAbs\ is equivalent with respect to $\equivE$ (structural
equivalence plus the eta law -- see Definition~\ref{d:etaeq})
$\abs x {\amb m {M .\nil}}$.\Qed

\begin{lem}
\label{l:TestCap}
 $\satDS P \TestCap$ iff $P$ is a simple ambient abstraction.
\end{lem}  

\proof
  We observe that $P$ has to be an I/O, and cannot be a message
  (otherwise by adding $\abs x \nil$ in parallel with $P$ we could
  violate the definition of \TestCap).
  
  Hence $P$ is an abstraction, and there are $M,N$ such that
\[
\begin{array}{rcl}
\msg M | P & \models & \diamond \amb m {\pAth{\ina n}} \andd  \Box
(\Dcomm \orr \amb m {\Ucomp}) \\[\tkpPOPL]
\msg N | P & \models & \diamond \amb m {{\zero}} \andd  \Box
(\Dcomm \orr \amb m {\Ucomp}) \\
\end{array} 
\]
By Lemma~\ref{l:auamCAP}, $P$ must be an \ambSemiAbs\ (note that
$\zero$ implies $\Ucomp$). Now Lemma~\ref{l:auamCAPbis} shows that $P$
must be an \ambAbs, which by Lemma~\ref{l:auamSIMPLECAP} is simple.\Qed

\myframe{ \UmsgN M \defi \Ucomm \andd \TestCap \limp \diamond \amb m
  {\pAth M} }

\begin{lem}
\label{l:UmsgCAP}
$\satDS P {\UmsgN M}$ iff $P \equiv \msg{M}$.\Qed
\end{lem}

\fi
\subsection{Synchronous Ambients}

\newcommand{\mayout}[1]{\may{\msg #1}}

Since the modal logic does not talk about the I/O primitives, it is
interesting to examine variations of these primitives, to see the
effect on the equality induced by the logic.  In MA communication is
asynchronous: since a message has no continuation, no process is
blocked until the message is consumed.  The most natural variation
consists in making communication \emph{synchronous}. For this the
production $\msg \eta $ for messages in the grammar of MA in
Table~\ref{ta:syn} is replaced by the production $\msg \eta . P$.
 Reduction
rule \trans{Red-Com} becomes:
\[
\shortaxiomC
{ \msg M .Q |
\abs x  P \longrightarrow 
Q| P \sub M x
} 
{Red-Com}
\]
The communication act liberates, at the same time, both the
continuation $P$ of the abstraction \emph{and} the continuation $Q$ of the
message.
We write \MAsync\ for the resulting synchronous calculus.

Synchrony leads to some important modifications in the assertions and
in the proofs of the results in the paper.  In \MAsync, the eta law
fails in the sense that the logic can separate eta equivalent terms
 (cf. Definition~\ref{d:etaeq}).
 Indeed, we will define a formula $\may{\msg
  n}.\A$ whose models are processes $\msg n.P$ with $P\Rar \sat\A$.
Then, returning to the eta law, formula $\mathtt{1input}\,\land\,
(\may{\msg n}.\amb{n}{\zero})\brtr \always\neg\mathtt{3Comp}$
is satisfied by $(x)\big(\msg x|(y)\nil)$, and not by
$(x)\nil$, where by $\mathtt{3Comp}$ we mean the formula $\mathtt{1Comp}|\mathtt{1Comp}|\mathtt{1Comp}$.

We will focus now on the characterisation of this new form of communication.
In asynchronous MA,  
our separation of  messages from abstractions 
exploited their asymmetry:  abstractions,  but not
messages,  have a continuation. In the synchronous case the asymmetry 
disappears,
therefore we have to use  a different route for the proof, which makes it a 
bit more involved.

Again, the most  delicate point is to find a replacement for the 
formula $\TestComm$. We  sketch how the new definition is
obtained.

\ifpopl
\else
\begin{itemize}
\item
\fi We first define a formula, $\OnlyCom$, that is satisfied  only by 
  abstractions $\abs x P$  and messages $\msg M . P$
 in which capability prefixes and
  ambients do  not appear in  the continuation $P$ and, moreover, no
  sub-term of  $P$
   contains more than two non-trivial  parallel
   components\iflong.\else:
\myframeC{ \OnlyCom\defi 
\begin{array}[t]{l}
\Ucomm \,\limpE\,  ( 
\begin{array}[t]{l}
 \Box (\Dcomm \orr \zero) \\[\tkpPOPL]
 \andd \; \diamond \zero)
\end{array}
 \end{array}
}
\fi 
 
\ifpopl
\else
\item
\fi
 Using $\OnlyCom$ we  define a formula, $\ComAmb$, that
is satisfied only by processes defined as those that satisfy
$\OnlyCom$ 
except that
  the innermost operator is an  ambient
\iflong
 $\amb \eta {\amb x \nil}$.
 \else  $\amb \eta {\amb y \nil}$:
\myframeC{
\begin{array}{l}
\ComAmb\defi  \\[\tkpPOPL]
\quad \thereis x \Big(
\begin{array}[t]{l}
\OnlyCom \,\limpE\,  \Big(
\begin{array}[t]{l}
\Box ( \begin{array}[t]{l}
\Dcomm 
\; \orr \; \amb x {\amb n \zero})
\end{array} \\[\tkpPOPL]
 \andd \;
\diamond \amb x {
\amb
n \zero}\Big)\end{array}  \\[\tkpPOPL]
\andd \;  \OnlyCom \,\limpE\,  \diamond \amb x {\amb  m \zero}  
\Big)
\end{array}
\\[\tkpPOPL]
\mbox{where $n$ and $m$ are different names.}
\end{array}
 }
\fi

\ifpopl
\else
\item
\fi
   We then define a formula that characterises the abstraction
  $\abs x {\amb  h{\amb x \nil}}$; we write 
$\Tcomm$ for $\Ucomm| \Ucomm| \Ucomm$:
\myframeC{
\begin{array}{l}
 \ImmN h \defi 
\begin{array}[t]{l}
\ComAmb    \\[\tkpPOPL]
\andd \;
\OnlyCom \limp  (\Box \myneg \Ucomm \: \andd \Box \myneg \Tcomm)
\\[\tkpPOPL]
 \andd \;
\OnlyCom \,\limpE\, \diamond \amb h {\amb n \true}  
\\[\tkpPOPL]
 \andd \;
\OnlyCom \,\limpE\, \diamond \amb h {\amb m \true}\,,  
\end{array}
\\[\tkpPOPL]
\multicolumn{1}{l}{ \mbox{where $n$ and $m$ are different names.}}
\end{array}
 }
Roughly, the first $\andd$-component implies   that a process that satisfies 
$\ImmN h$ has an abstraction or a message as its outermost operator, and
an ambient 
$ {\amb  \eta{\amb x \nil}}$
 as the innermost. The second
$\andd$-component, call it $\FFF$, 
ensures that the process does not have any other operators; that is,
the ambient
$ {\amb  \eta{\amb x \nil}}$
is reached
immediately after the  initial communication.
For instance, the process $R \defi \msg M . \abs x {\amb  h{\amb x \nil}}$
 does 
not satisfy $\FFF$ because $R |\abs x \nil \Longrightarrow
\abs x { \amb  h{\amb x \nil} }
$ 
and $\abs x {\amb  h{\amb x \nil}}$  satisfies $\Ucomm$.
Finally, the third and fourth $\andd$-components  rule out the messages and 
 the abstraction $\abs x {\amb x
  {\amb x \nil}}$.
\ifpopl
\else
\end{itemize} 
\fi

Once we have defined formulas to capture primitives for synchronous
communication, the other expressiveness results in the paper
also hold for synchronous \MA. The corresponding proofs follow closely
the arguments in the previous sections.


We now move to the formal definition and analysis of the formulas we
alluded to above.



\subsubsection*{Modifications between Lemma~\ref{l:mauam} 
and \ref{l:UcommNN}}
             
To define a formula that captures synchronous outputs
(Lemma~\ref{l:synMsgA} below), we introduce tester processes of the
form $\abs x{\amb h{\amb x\nil}}$, for a given name $h$. The logical
characterisation of these (Lemma~\ref{l:synImm}) is slightly more
complicated than the corresponding result in the asynchronous case
(Lemma~\ref{l:TestComm}), and is based on four grammars describing
communicating processes, that are defined as follows.

\myframe{ \OnlyCom\defi 
\begin{array}[t]{l}
\Ucomm \,\limpE\,  (\Box (\Dcomm \orr \zero) \: \andd \diamond \zero)
\end{array}
}
\ifREADERS\finish{sopra, ho tolto $ \andd  \Ucomm$, non credo serva } 
\fi

To interpret formula \OnlyCom, we introduce the following grammars:
\[\eqalign{
  H :: =
& \msg \eta . \nil \midd \abs x \nil \midd   \msg \eta . (H|H) \midd 
\abs x {(H|H) } \midd \msg \eta . H \midd \abs x H \cr
  \HAM :: =
& \msg {\eta'} .\amb \eta{ \amb  y \zero} \midd 
\abs x {\amb \eta{ \amb  y \zero}} \midd
   \msg {\eta'} . (H | \HAM) \midd 
\abs x {(H | \HAM) } \midd \msg {\eta'} \HAM \midd \abs x \HAM \cr
  H^\star :: =
& H \midd  \nil  \cr
  K^\star :: =
&  \msg {\eta'} .\amb \eta{ \amb  {\eta''} \zero} \midd 
\abs x {\amb \eta{ \amb  {\eta''} \zero}} \midd
   \msg {\eta'} . (H | \HAM) \midd 
\abs x {(H | \HAM) } \midd \msg {\eta'} \HAM \midd \abs x \HAM \midd 
\amb\eta{\amb {\eta'}{\nil}}
}
\]

We write 
\begin{itemize}
\item
$ \GRH$ for the set of terms described by $H$, 

\item  $ \GRHAM$ for those described by $K$, 

\item  $\GRHstar$ for those described by $H^\star$, and 

\item  $\GRHAMstar$ for those described by $K^\star$.
\end{itemize}
(The grammar for $K^\star$, with respect to  that for $K$, has the additional
production for $ \amb\eta{\amb {\eta'}{\nil}}$, and has $
\amb\eta{\amb {\eta'}{\nil}}$ in place of $ \amb\eta{\amb {y}{\nil}}$
in the other productions.)

\begin{lem}
\label{l:synOnlyComSUB}
Suppose $P $ is a \MAsync\ process.
$P \sub n x\in \GRH$  iff $P \in \GRH$.\Qed
\end{lem} 

\begin{lem}
\label{l:synOnlyCom}
Suppose $P $ is a \MAsync\ process and $\satDS P \OnlyCom$. Then $P \equiv P'$ for some $P'  \in \GRH$. 
\end{lem} 

\proof
Suppose $\satDS {P_1} {\OnlyCom}$. Then there is  a process $P_2$ with
$\satDS{P_2}\Ucomm$ such that 
$$ \satDS{P_1|P_2}{(\Box (\Dcomm \orr \zero)
  \: \andd \diamond \zero)}.$$ 
In particular, it holds that $ \satDS{P_1|P_2}{\Dcomm}$, hence  
$ \satDS{P_1}{\Ucomm}$.

We show that if $Q_1, Q_2$ are processes that satisfy $\Ucomm$ and such 
that  $$ \satDS{Q_1|Q_2}{(\Box (\Dcomm \orr \zero)
  \: \andd \diamond \zero)}, $$ 
then
$Q_1,Q_2 \in\GRH$.

The proof is by induction on the maximal depth of $Q_1,Q_2$.  The case
when this depth is $1$ is easy.  If this depth is greater than $1$,
then $Q_1 | Q_2 \longrightarrow Q'_1 | Q'_2$, using the \trans{com}
rule, where $Q_i$ is the continuation of $Q_i$.  We have three cases:
\begin{itemize}
\item $Q'_1 \equiv \nil$, $Q_2' \equiv R_1|R_2$ for some non-trivial
  $R_1,R_2$; 
\item the symmetric case;
  
\item none of $Q_1'$ and $ Q_2'$ is structurally congruent to $\nil$.
\end{itemize}  
In the first two cases, we deduce that $R_1,R_2$ satisfy $\Ucomm$,
and then use induction to infer $R_1,R_2 \in \GRH$.
Then using the first 4 productions of the grammar, and 
Lemma~\ref{l:synOnlyComSUB}, 
$Q_1,Q_2\in \GRH$.
 In the third case, 
use induction to infer $Q'_1,Q'_2 \in \GRH$.  Hence  also 
$Q_1,Q_2 \in \GRH$, using the last 2 productions of the grammar and 
Lemma~\ref{l:synOnlyComSUB}.\Qed

\begin{lem}
\label{l:synComAmbAUX}
Suppose $P $ is a \MAsync\ process, and 
$\satDS{P}{ \Box (\Dcomm \orr \amb h {\amb n \zero}) \: \andd
\diamond \amb h {
\amb
n \zero} }$, where $h,n$ are any names. Then $P \equiv P_1 |P_2$ with 
$P_1 \in \GRH$ and $P_2 \in \GRHAMstar$.
\end{lem} 

\proof
By induction on the size of $P$, where the size is the number of
operators in $P$.
 The size cannot be $0$ or $1$. If the size is $2$
then $P =  \amb h {\amb n \zero}$, and $P \equiv \nil | \amb h {\amb n
  \zero}$, hence the assertion of the lemma, for $P_1\defi \nil$. 

Suppose the size if greater than $2$.
Call 
\[ \FFF ~\defi~  \Box (\Dcomm \orr \amb h {\amb n \zero}) \: \andd
\diamond \amb h {
\amb
n \zero}\]

 Then 
\[ P \equiv P_1 | P_2 \quad 
\mbox{ where, for $i=1,2$,we have } \quad \satDS {P_i } \Ucomm\] 
Since $P$ must reduce, 
\[ P_1 | P_2 \,\equiv\, \msg m . Q_1 | \abs x {Q_2} \]
and 
\[ \satDS{Q_1 | Q_2 \sub m x}{\FFF}\,.\]
The size of $Q_1 | Q_2\sub m x $ is smaller.

It can be that $Q_1$ or $Q_2$ are $\nil$, or none is $\nil$ (they
cannot both be $\nil$). In both cases we can conclude by referring to
the appropriate grammar productions and by using the inductive
hypothesis.\Qed

Define now:
\myframe{
\begin{array}{l}
\ComAmb\defi 
\thereis x \Big(
\begin{array}[t]{l}
\OnlyCom \,\limpE\,  \Big(\Box (\Dcomm \orr \amb x {\amb n \zero}) \: \andd
\diamond \amb x {
\amb
n \zero}\Big) \\[\tkpPOPL]
\andd\quad  \OnlyCom \,\limpE\,  \diamond \amb x {\amb  m \zero}  
\Big)
\end{array}
\\[\tkpPOPL]
\mbox{where $n$ and $m$ are different names.}
\end{array}
 }
\begin{lem}
\label{l:synComAmb}
Suppose $P $ is a \MAsync\ process, and 
 $\satDS P \ComAmb$. Then $P \equiv P'$ for some $P'  \in \GRHAM$. 
\end{lem} 

\proof
Suppose $\satDS {P_1} {\ComAmb}$. Then there is  a process $P_2$ and
some $h$ with
$\satDS{P_2}\OnlyCom$ such that $ \satDS{P_1|P_2}
\Box (\Dcomm \orr \amb h {\amb n \zero}) \: \andd
\diamond \amb h {
\amb
n \zero}$. By Lemma~\ref{l:synOnlyCom}, 
$P_2 \equiv  \in \GRH$. 
By Lemma~\ref{l:synComAmbAUX}, 
$P_1 \in \GRHAMstar$.
Moreover, since $\satDS{P_1}\Ucomm$, it holds that
$P_1 \not \equiv \amb h{\amb n \nil}$; from this and 
$\satDS {P_1} {\OnlyCom \,\limpE\,  \diamond \amb h {\amb  m \zero}}$, we
deduce  
$P_1 \in \GRHAM$.



Define 
\myframe{
\begin{array}{l}
 \ImmN h \defi 
\begin{array}[t]{l}
\ComAmb \: \andd   \\[\tkpPOPL]
\OnlyCom \limp  (\Box \myneg \Ucomm \: \andd \Box \myneg \Tcomm)
\\[\tkpPOPL]
\OnlyCom \,\limpE\, \diamond \amb h {\amb n \true}  
\\[\tkpPOPL]
\OnlyCom \,\limpE\, \diamond \amb h {\amb m \true}  
\end{array}
\\[\tkpPOPL]
\multicolumn{1}{l}{ \mbox{where $m \neq n$.}}
\end{array}
 }
\begin{lem}
\label{l:synImm}
Suppose $\satDS P \ImmN h$. Then $P \equiv 
\abs x{\amb   h {\amb x\nil}}$. 
\end{lem} 

\proof By Lemma~\ref{l:synComAmb}, $P \equiv P' \in
 \GRHAM$.
One then shows that $\abs x{\amb h {
\amb  x \nil}}$ satisfies $\ImmN h$, whereas the other terms in $\GRHAM$ 
do not (choosing the
appropriate $\OnlyCom$).\Qed

Now we can define the formula:
\myframe{\mayout n . \A \defi
\begin{array}[t]{l}
 \Ucomm \\[\tkpPOPL]
\andd\;  \all x ( \ImmN x  \limp \diamond( \amb x {\amb n\zero} | \A))
\end{array}
  }

\begin{lem}
\label{l:synMsgA}
Suppose $P $ is a \MAsync\ process.
It holds that $\satDS P {\mayout n . \A}$ iff $P \equiv \msg n . Q$ and $Q 
\Longrightarrow Q'$ and 
$\satDS{Q'}\A$.
\end{lem} 

\proof
Take $h$ fresh. Then 
by Lemma~\ref{l:synImm}, 
\[ 
\satDS{ P | \abs x {\amb h {\amb x \nil}}}{ \diamond 
(\amb
  h{\amb x \nil} |\A)}\,.
\]
From this, and $\satDS P \Ucomm$, we deduce 
$P \equiv \msg m . P'$, for some $m,P'$. We also deduce that
\[ 
\satDS{ P' | {\amb h {\amb m \nil}}}{ \diamond (\amb
  h{\amb m \nil} |\A)}\,.
\]
Since $h$ is fresh, $P'$ cannot interact with $h$. Hence  $m =n$, and
moreover $P' \Longrightarrow  \models \A$.\Qed

\begin{lem}
\label{l:UabsM}
Suppose $P $ is a \MAsync\ process.
It holds that $\satDS P{\UabsN M . A}$ iff $P \equiv \abs x {P'}$ 
and $P' \sub n x \Longrightarrow P'' \models A$.\Qed
\end{lem}  

\subsection{Other Extensions}

\subsubsection{Name restriction and revelation}
Usually~\cite{CaGo98,CaGo99popl}, the syntax of MA also has the
restriction operator.
In~\cite{Cardelli::Gordon::NameRestriction::01}, Cardelli and Gordon
propose an extension of AL with logical connectives to describe
restriction.  In particular, the operator of \emph{name revelation}
allows one to derive $\copyright n$
(Subsection~\ref{subsec:copyright}). In presence of restriction in the
calculus, we cannot adapt our construction to capture finiteness of
processes, intuitively because our approach consists in exhibiting a
context that allows a finite process to reduce to $\nil$, which is
not possible in general in presence of restriction.
However, characteristic formulas can be derived, by
enriching our constructions with a formula that says that a process
has no restriction (which is definable using name revelation).

\subsubsection{Strong sometimes modality}

One could consider a ``strong'' version of the sometimes ($\sometime$)
 modality,
where $\longrightarrow$ replaces $\Rar$ in the definition of $\sat$.
This variant is  easier to study, and less interesting
in a sense. We explain the effects it would have. 
The only drawback is that
with  a
strong version of $\sometime$ we could not derive
the formulas of Section \ref{s:intensional}, and as a
consequence characteristic formulas can be given for finite processes
only. On the other hand, the
 formulas for capabilities and communications would become
much simpler; we would not have to consider stuttering and eta
conversions; logical equivalence would coincide with structural
congruence.

\subsubsection{Recursion}
In a different direction, a variant of \MA\ can be considered in which
a recursion operator is used instead of replication (see for
example~\cite{LeSaTOPLAS}). Recursion gives trees with infinite depth;
this prevents us from defining the measures $\ds(P)$ and $\dd(P)$ up
to structural congruence.  Moreover, the constructions in
Subsection~\ref{subsec:persistence} are based on the characterisation
of persistence (that provides a form of `recursion in width') of
replicated processes. We do not think that they could be easily
adapted to a calculus with recursion.

\subsubsection*{Acknowledgments}

We  thank the anonymous referees for their careful reading of the
paper,  and for their comments and suggestions, which resulted in a number
of improvements for the paper.



\bibliographystyle{alpha}
\bibliography{DSbib,OTHERSbib,refs}

\ifREADERS
\newpage
\input{todo}
\newpage
\input{notes}
\fi
\end{document}